\documentclass[aps, prd, preprintnumbers, floatfix, showpacs, showkeys, nofootinbib, 10pt]{revtex4-1}

\usepackage[dvips]{graphics}
\usepackage{amsmath,amsfonts,bm}
\usepackage{amssymb}
\usepackage{amsthm}
\usepackage{epsfig,bm}
\usepackage{feynmp}
\usepackage{graphicx,comment}
\usepackage{dcolumn,color}
\usepackage{mathtools}
\usepackage{fancyhdr}
\usepackage{bm}
\usepackage{cancel}
\usepackage{slashed}
\usepackage{verbatim}
\usepackage{listings}
\usepackage{graphicx,latexsym}
\usepackage{rotating}
\usepackage{color}
\usepackage{float}
\usepackage{enumerate}
\usepackage{array}
\usepackage{tabularx}
\usepackage{longtable}
\usepackage{booktabs}

 \def\cB{{\cal B}}

 \def\cR{{\cal R}}  
   
 \def\cR{{\cal R}}
\def\tr{\mathop{\rm Tr}}  \def\Tr{\mathop{\rm Tr}}

                 \def\d{\partial}

\def\beq{\begin{equation}}
\def\eeq{\end{equation}}
\def\bea{\begin{eqnarray}}
\def\eea{\end{eqnarray}}

\newcommand{\be}{\begin{equation}}
\newcommand{\ee}{\end{equation}}
\newcommand{\bear}{\begin{eqnarray}}
\newcommand{\eear}{\end{eqnarray}}
\newcommand{\ba}{\begin{array}}
\newcommand{\ea}{\end{array}}

\newcommand\tT{\widetilde{T}}

\newcommand{\of}[1]{\!\left(#1\right)}
\newcommand{\sqof}[1]{\left[#1\right]}

\newcommand{\RET}{\nonumber \\}

\begin{document}
\preprint{NYU-6-26-2014,  EFI-14-16}
\preprint{\today}


\title{Central Production of $\eta$ and $\eta'$ via Double Pomeron Exchange in the Sakai-Sugimoto Model}


\author{Neil~Anderson$^1$, Sophia~K.~Domokos$^2$, Jeffrey~A.~Harvey$^3$, Nelia~Mann$^1$}
\affiliation{$^1$  Department of Physics,
Reed College, Portland OR 97202, USA}
\affiliation{$^2$ Center for Cosmology and Particle Physics, Department of Physics,
New York University, New York NY 10003, USA, }
\affiliation{$^3$ Enrico Fermi Institute and Department of Physics, University of Chicago, Chicago IL 60637, USA}


\begin{abstract}

We construct a string-inspired model for the central production of $\eta$ and $\eta'$ mesons in proton-proton collisions via double Pomeron exchange.  Using general symmetry considerations, we construct a low-energy differential cross section for double glueball exchange in terms of some undetermined coupling constants and form factors.  We extend this model to the Regge regime, replacing the glueball propagators with Pomeron trajectories, and modifying the interaction term by a factor derived from the 5-string scattering amplitude in flat space.  We then fix the couplings which remain undetermined, using the Sakai-Sugimoto framework to model  low-energy QCD. Finally, we generate a simulation of the scattering process at $\sqrt{s} = 29.1$ GeV, where double Pomeron exchange should play a role (secondary to double Reggeon exchange).  We focus on the dependence of the scattering cross section on $\theta_{34}$, the angle between the scattered protons in the transverse plane.  The results exhibit a definite deviation from the pure $\sin^2\theta_{34}$ dependence that arises as a consequence of natural parity violation alone.  The amount of deviation is primarily determined by couplings that come from the Chern-Simons action of the AdS/QCD supergravity dual, which is directly related to the QCD gravitational anomaly, and thus constitutes a universal part of any five-dimensional string/gravity dual theory of QCD.  We argue that this makes the high energy central production of pseudoscalar mesons an interesting probe of AdS/QCD models.
\end{abstract}



\maketitle


\section{Introduction}

Over the past ten years,  gauge-string duality has found fruitful application in the realm of strongly coupled systems, from low-energy QCD to superconductors. The connection between string theory and QCD is far older, however. It dates back to the days of Regge theory, when string theories were constructed in order to describe hadronic spectra, and hadronic scattering at high center-of-mass energy $s$ and small momentum transfer $t$ (the Regge regime) \cite{Collinsbook, Veneziano}. To good approximation, the masses of mesons and baryons lie on linear trajectories -- Regge trajectories -- with particle spin $J$ related to the mass $m$ as $J = \alpha_0 + \alpha' m^2$.  String theories generate precisely such a spectrum. Meanwhile, scattering events at small $t$ should be governed by the exchange of hadrons. In the Regge regime, hadrons of arbitrarily high spin contribute to the process, so the entire Regge trajectory should be taken into account. The mediating hadron can be replaced by a ``Reggeon,'' which couples like the lowest state on the exchanged trajectory. The Veneziano amplitude, now known to describe the scattering of strings in flat space, was first proposed as a phenomenological model for the scattering of hadrons in the Regge limit.

The old idea of  treating Reggeons as flat space strings captures many qualitative features, but fails to accurately describe the observed Regge trajectories and scattering processes.  It may yet find new life in gauge-string duality, whereby certain gauge theories (like QCD) are mapped onto string theories in higher dimensional curved spacetimes. Regge regime hadronic scattering  can thus be translated to the holographic dual, where hadronic scattering is quite literally string scattering -- but in a curved 5d space.

Consider, for example, the behavior of proton-proton or proton-anti-proton scattering in the Regge regime.  Since the cross-sections behave similarly at very large $s$,  the exchanged object should be insensitive to the charges of the scattered (anti)protons \footnote{At smaller $s$, the exchange of mesonic trajectories also contributes.}. The exchanged trajectory thus has vacuum quantum numbers, and is known as the {\it Pomeron}. The lightest particle on the Pomeron trajectory is widely believed to be a $J^{PC}= 2^{++}$ glueball  \footnote{Though there are some arguments in the literature  that it may in fact be a vector particle \cite{CGM, Cudell, bfkl, DLbook}.}.  Since the Pomeron consists of even spin glueballs, its holographic dual should be a closed string \cite{bpst, Janik}. The Pomeron was first identified in a holographic context in \cite{Brower:2006ea} as the Regge trajectory of string states in an asymptotically Anti-de Sitter (aAdS) space, whose lowest state is the graviton.  This means that hadron scattering mediated by Pomeron exchange should be equivalent to closed string scattering in the holographic dual.  Though such amplitudes are difficult to calculate exactly, one can model the interactions of the lowest modes on the Regge trajectories in the supergravity limit of gauge-string models, and then extend the results for these lowest-energy states to the Regge regime.

Although no known holographic dual model perfectly reproduces all relevant features of low-energy QCD, several reasonable toy models exist. The Sakai-Sugimoto model, for instance, consists of $N_f$ D8 and $\overline{\mathrm{D8}}$ branes in the warped gravitational background generated by $N_c$ D4-branes \cite{Sakai:2004cn} (for large $N_c$).  The open strings on the D-branes are dual to mesons, while the closed strings living in the bulk are dual to glueballs.  Despite some important limitations, the masses of mesons and some glueballs computed in this framework have been found to match experimental and lattice results with reasonable accuracy \cite{Sakai:2005yt, Brower:2000rp, Meyer:2004jc}.

It is interesting and important, then, to extend the results of low-energy holographic duals to modeling the scattering processes of baryons and mesons in the Regge limit.  Glueball scattering (in the Regge limit and beyond) was studied in \cite{Polchinski:2001tt}.  Meanwhile, \cite{Domokos:2009hm} proposed a more phenomenological approach  for studying proton-(anti)proton scattering holographically, relying on the assumption that string scattering in weakly curved backgrounds should roughly take the same form as \textit{flat space} string scattering.  First, \cite{Domokos:2009hm}  computed the amplitude for holographic proton-proton scattering via spin 2 glueball exchange from the supergravity limit of the Sakai-Sugimoto model. They then used the structure of the flat-space Virasoro-Shapiro amplitude to model the full Pomeron propagator, and substituted this propagator for the glueball propagators in the proton-proton scattering amplitude. We call this procedure``Reggeizing"  the amplitude -- in other words, extending the low energy result to the Regge regime. The slope and intercept of the Pomeron trajectory were left as free parameters in this procedure, as these should differ from their flat space values. The result of \cite{Domokos:2009hm}'s procedure was a phenomenological model for high energy scattering which could be directly compared with data.

In this work, we use similar techniques to model the double-Pomeron-mediated central production of $\eta$ and $\eta'$ mesons in Regge-regime proton-proton scattering.  We begin by determining the differential cross section for the central production of a pseudoscalar meson via the exchange of spin 2 particles.  This process violates natural parity. In AdS/QCD, natural parity violating couplings arise from bulk Chern-Simons terms present in all D-brane constructions which yield QCD in the low-energy limit. These terms are responsible for reproducing the gravitational (and chiral) anomalies of QCD\footnote{Recall that gauge-gravity duality equates the classical supergravity theory in curved space with the fully quantum flat space field theory (at large 't Hooft coupling and large $N_C$).}. Since the coefficients of these terms are fully fixed -- from supergravity and field theory -- their inclusion in an AdS/QCD model does not increase the number of free parameters, and is furthermore relatively model-independent.

After computing the low energy amplitude, we  use the form of a five (closed) string amplitude to motivate the ``Reggeization'' of the process.  This does not lead to the same prediction as separately Reggeizing each glueball propagator, as was done in \cite{KMV}.  Our treatment is based on an analysis of central production of particles by double Pomeron exchange in the string dual \cite{Herzog, BDT}, where it was demonstrated that the behavior of the scattering amplitude depends on the mass of the centrally produced particle. We will see that the relatively light mass of the pseudoscalar meson we are considering strongly affects how the glueball propagators should be Reggeized.  We finish by generating a Monte Carlo simulation of $\eta/\eta'$ central production at $\sqrt{s} = 29.1$ GeV.  We choose this energy to facilitate comparison with the WA-102 experiment \cite{WA102}. It is important to note that double Pomeron exchange only accounts for a small part of $\eta$ and $\eta'$ production in this regime, with significant contributions coming from double Reggeon exchange. However, double Pomeron exchange dominates at much higher energies, and there our model, as it stands, should represent a reasonable approximation.

The central ingredients of our construction yield clear experimental signatures to distinguish them from other approaches, thus providing a powerful check of its underlying principles. First of all, treating the Pomeron as a trajectory of even spin glueballs implies a particular class of couplings, which would not be present if the lowest state were a vector-like particle. The effects of this distinction are apparent in the form of the differential cross sections.

In addition, because the Pomeron should be flavor neutral, the Pomeron-Pomeron-pseudoscalar interaction exclusively involves the flavor singlet pseudoscalar meson, which we will call $\eta_0$. The couplings to $\eta$ and $\eta'$ mesons are generated using the mixing angle with the flavor singlet. As pions can only be centrally produced by exchanging two mesonic Regge trajectories (``Reggeons''), the sole production of $\eta$ and $\eta'$ becomes a unique signature for double Pomeron processes.

Furthermore, the coefficient of the bulk gauge-gravity Chern-Simons term which generates our coupling is uniquely fixed by requiring consistency of the supergravity theory on one side of the duality, and by the gravitational anomaly on the other. This makes our predictions relatively model-independent, though explicit predictions for the glueball-glueball-meson couplings do require us to choose a particular holographic QCD dual.

Finally, the Reggeization procedure based on the 5-string amplitude has a clear experimental signature distinct from the naive Reggeization of individual graviton propagators.  All of this suggests that comparison to experiment, either by supplementing these calculations with those for double Reggeon exchange or by considering experiments run at higher center-of-mass energy, may yield significant new insights.\footnote{Measuring the exclusive production of pseudoscalar mesons at the multi-TeV scale would be very difficult, though perhaps not impossible.}

The body of this work begins in section II, where we calculate the cross section for central production of $\eta$ or $\eta'$ via $t$-channel glueball exchange in the Regge limit.  In section III we discuss the 5-string flat space amplitude and describe the Reggeization of our differential cross section.  In section IV we compute the necessary low-energy couplings  in the Sakai Sugimoto model. In section V, we present the results of a Monte Carlo simulation for our model.  Finally, we discuss the results and suggest some future work in section VI.  A detailed description of the kinematics and phase space of $ 2 \rightarrow 3$ scattering in the Regge limit is provided in the Appendix.

\section{The Feynman Amplitude and the Cross Section}

In the Regge limit, production of $\eta$ and $\eta'$ mesons in proton-proton scattering is dominated by processes involving the $t$-channel exchange of Pomerons.  We begin by reviewing the kinematics and phase space of $2 \rightarrow 3$ scattering, and the implications of the Regge limit for this process.  We then compute the  amplitude and cross section for producing  $\eta$ or $\eta'$ mesons via t-channel double glueball exchange in proton-proton scattering.  Because the glueball is flavor neutral, it only couples to the flavor singlet in the pseudoscalar meson nonet.  We therefore use the mixing angle to determine the relationship between $\eta$ and  $\eta'$ production.

\subsection{The Kinematics and Phase Space of $2 \rightarrow 3$ Scattering in the Regge Limit}

\begin{figure}
\begin{center}
\resizebox{1.9in}{!}{\includegraphics{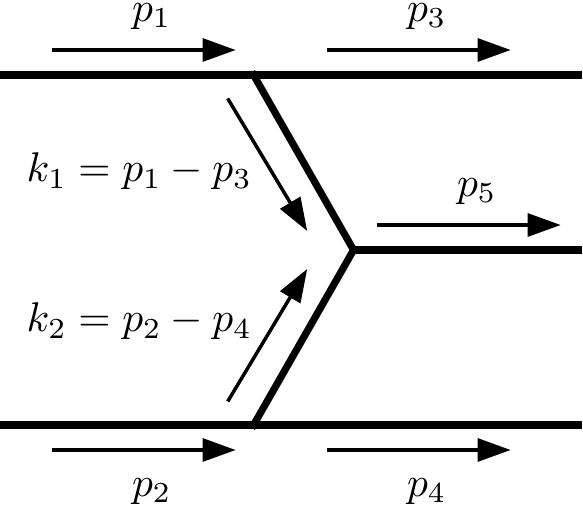}}
\caption{\label{kinematics} The kinematics of $2 \rightarrow 3$ central production.}
\end{center}
\end{figure}

Consider a $2 \rightarrow 3$ central production process: there are two incoming protons with momenta $p_1$ and $p_2$, two outgoing protons with momenta $p_3$ and $p_4$, and one outgoing pseudoscalar meson with momentum $p_5$, shown in figure \ref{kinematics}.  The mass-shell conditions (with a mostly plus metric) are
\beq
\label{eqn:massshell}
p_1^2 = p_2^2 = p_3^2 = p_4^2 = -m_p^2, \hspace{1in} p_5^2 = -m_5^2,
\eeq
where $m_p$ is the mass of the proton, and $m_5$ is the mass of either the $\eta$ or the $\eta'$ particle.
 In addition, conservation of four-momentum yields
\beq
\label{eqn:conservation}
p_1 + p_2 = p_3 + p_4 + p_5.
\eeq
To make calculations more convenient, we can define five Mandelstam variables,
\beq
s = -(p_1 + p_2)^2, \hspace{.4in} t_1 = -(p_1 - p_3)^2, \hspace{.4in} t_2 = -(p_2 - p_4)^2, \hspace{.4in} s_1 = -(p_3 + p_5)^2, \hspace{.4in} s_2 = -(p_4 + p_5)^2.
\eeq
We will assume the initial protons' momenta are equal and opposite, aligned along the $z$-axis, so that we can write the five four-momenta as
\beq
p_1 = (E, 0, 0, p), \hspace{.3in} p_2 = (E, 0, 0, -p), \hspace{.3in} p_3 = (E_3, {\bf q}_3, p_{3z}), \hspace{.3in} p_4 = (E_4, {\bf q}_4, p_{4z}), \hspace{.3in} p_5 = (E_5, {\bf q}_5, p_{5z}). \hspace{.2in}
\eeq
There are five independent kinematic variables determining the outgoing particles' momenta.  However, the azimuthal symmetry of the initial states ensures that the final states will only depend on four.  We will use $t_1$ and $t_2$ as two of these variables.  A third will be the angle $\theta_{34}$ between the transverse portions of momentum for the outgoing protons, defined as
\beq
{\bf q}_3 = (q_3\cos\theta_3, q_3\sin\theta_3), \hspace{.5in} {\bf q}_4 = (q_4\cos\theta_4, q_4\sin\theta_4), \hspace{.5in} \theta_{34} = \theta_4 - \theta_3.
\eeq
The fourth will be $x_F$, the difference between the fractions of initial longitudinal momentum carried by the outgoing protons, defined as
\beq
p_{3z} = x_1p, \hspace{1in} p_{4z} = -x_2p, \hspace{1in} x_F = x_1 - x_2.
\eeq
In the Regge limit, where the center-of-mass energy is large and the scattering angles of the two protons are small, we will have $s \gg s_1, s_2 \gg t_1, t_2, m^2$ (where $m$ could be the mass of any of the particles involved).  We also assume that the quantity
\beq
\label{eqn:eta}
\mu = \frac{s_1 s_2}{s},
\eeq
remains fixed in the Regge limit, and of magnitude comparable to $t_1, t_2$ or $m^2$ for any of the involved masses\footnote{In the literature this parameter is generally known as $\eta$; we have renamed it to avoid confusion with the $\eta$ and $\eta'$ mesons.}.  In Appendix A, we analyze the kinematics and phase space in the Regge limit.  We find that the phase space is dominated by the region near $x_F = 0$, and that we can write the total cross section as
\beq
\label{eqn:cross}
\sigma \approx \frac{1}{4(4\pi)^4 s^2} \int \langle |\mathcal{A}|^2\rangle \, \ln\left(\frac{s}{\mu}\right) \, d\theta_{34} \, dt_1 \, dt_2,
\eeq
where the spin-averaged amplitude squared $\langle |\mathcal{A}|^2\rangle$ is evaluated in the Regge limit at $x_F = 0$.  In this limit, we also have
\beq
\mu \approx m_5^2 - t_1 - t_2 + 2\sqrt{t_1 t_2} \, \cos\theta_{34}, \hspace{.5in} s_1 \approx s_2 \approx \sqrt{s\mu},\hspace{.5in}  q_3 \approx \sqrt{-t_1}, \hspace{.5in} q_4 \approx \sqrt{-t_2}.
\eeq

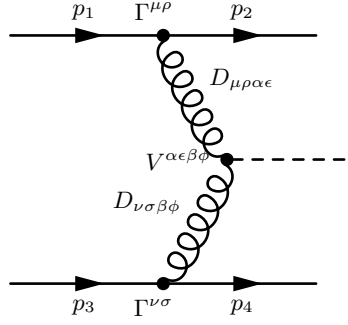
\begin{figure}
\begin{center}
\vspace{.4in}
\begin{fmffile}{graph1}
\begin{fmfgraph*}(2,1.3)
\fmfleft{i1,i2}
\fmfright{o1,o2,o3}
\fmf{fermion,label.side=right,label=$p_3$}{i1,v1}
\fmf{fermion,label.side=right,label=$p_4$}{v1,o1}
\fmf{fermion,label.side=left,label=$p_1$}{i2,v2}
\fmf{fermion,label.side=left,label=$p_2$}{v2,o3}
\fmffreeze
\fmf{curly,label=$D_{\mu\rho\alpha\epsilon}$}{v2,v3}
\fmf{curly,label.side=left,label=$D_{\nu\sigma\beta\phi}$}{v1,v3}
\fmf{dashes}{v3,o2}
\fmfdot{v1,v2,v3}
\fmfv{label=$\Gamma^{\mu\rho}$\hspace{.1in}}{v2}
\fmfv{label=$\Gamma^{\nu\sigma}$\hspace{.1in}}{v1}
\fmfv{label.angle=180,label=$V^{\alpha\epsilon\beta\phi}$}{v3}
\end{fmfgraph*}
\end{fmffile}
\vspace{.4in}
\caption{\label{feynman} The Feynman amplitude for the central production process. }
\end{center}
\end{figure}

\subsection{The Feynman Amplitude}
The Feynman diagram we need to compute is shown in figure \ref{feynman}.  It involves the propagator for the glueball,
a massive spin 2 object (see e.g. \cite{prop}):
\beq
D_{\mu\rho\nu\sigma}(k) =  \frac{id_{\mu\rho\nu\sigma}(k)}{k^2 + m_g^2},
\eeq
with

\beq
d_{\mu\rho\nu\sigma}(k) = \frac{1}{2}\left(\eta_{\mu\nu}\eta_{\rho\sigma} + \eta_{\mu\sigma}\eta_{\rho\nu}\right) + \frac{1}{2m_g^2}\left(k_{\mu}k_{\sigma}\eta_{\rho\nu} + k_{\mu}k_{\nu}\eta_{\rho\sigma} + k_{\rho}k_{\sigma}\eta_{\mu\nu} + k_{\rho}k_{\nu}\eta_{\mu\sigma}\right)
\eeq
$$
+ \frac{1}{24}\left[\left(\frac{k^2}{m_g^2}\right)^2 + 3\left(\frac{k^2}{m_g^2}\right) - 6\right]\eta_{\mu\rho}\eta_{\nu\sigma} - \frac{(k^2 + 3m_g^2)}{6m_g^4}\left(k_{\mu}k_{\rho}\eta_{\nu\sigma} + k_{\nu}k_{\sigma}\eta_{\mu\rho}\right) + \frac{2k_{\mu}k_{\rho}k_{\nu}k_{\sigma}}{3m_g^4},
$$
where $k$ could be either $k_1$ or $k_2$, and $m_g$ is the mass of the glueball.  We will see that due to the structure of the vertices, only the first term in this expression contributes.  We also need two vertices: a proton-proton-glueball vertex and a glueball-glueball-pseudoscalar vertex.

We assume that the glueball couples primarily to the stress-energy tensor of the protons, as inspired by the idea of tensor meson dominance \cite{freundtmd, tmdII}.  In this case, the proton-proton-glueball vertex can be written as
\beq
\Gamma^{\mu\rho} = \lambda_{\mathcal{P}}\left[\frac{A(t)}{2}\left(\gamma^{\mu}P^{\rho} + \gamma^{\rho}P^{\mu}\right) + \frac{B(t)}{8m_p}\Big(P^{\mu}[\gamma^{\rho}, \gamma^{\nu}] + P^{\rho}[\gamma^{\mu},\gamma^{\nu}]\Big)k_{\nu} - \frac{C(t)}{m_p}\left(\eta^{\mu\rho}t + k^{\mu}k^{\rho}\right)\right],
\eeq
with $k = p - q$, and $P = (p + q)/2$.  The form factors $A,B,C$ are then derived from the energy-momentum tensor, which implies that $A(0) = 1$ and $B(0) = 0$ \cite{Pagels}.  In our amplitude, the term proportional to $C(t)$ will not contribute: it vanishes when contracted through the glueball propagator with the natural-parity-violating central vertex.  We also drop terms proportional to $B(t)$, as $B(t)$ is small and slowly varying near $t = 0$, where the amplitude is largest.  This is supported by calculations in the Sakai-Sugimoto model, as discussed in \cite{Domokos:2009hm}.

Meanwhile, the glueball-glueball-pseudoscalar vertex is
\beq
\label{eq:vertex}
V^{\alpha\epsilon\beta\phi} = \left[G_1(t_1, t_2)\eta^{\epsilon\phi} + G_2(t_1, t_2)k_2^{\epsilon}k_1^{\phi}\right]\varepsilon^{\alpha\beta\gamma\delta}k_{1\gamma}k_{2\delta},
\eeq
with
\beq
k_1 = p_1 - p_3, \hspace{.5in} k_2 = p_2 - p_4.
\eeq
The structures appearing here are the only ones allowed by the symmetries of the strong force: parity and charge conjugation in addition to Lorentz symmetry.  Note the presence of the natural parity violating epsilon tensor, which on the gravity side of the duality arises from the Chern-Simons interaction.  We will eventually take the as-yet-arbitrary factors  $G_1$ and $G_2$
from the Sakai-Sugimoto model, along with the coupling constant $\lambda_{\mathcal{P}}$ and the form factor $A(t)$ of the proton-proton-glueball vertex.

The full amplitude for central production of pseudoscalars in double glueball exchange processes is thus
\begin{align}
\mathcal{A} = \Big(\bar{u}_3\Gamma^{\mu\rho}u_1\Big)D_{\mu\rho\alpha\epsilon}V^{\alpha\epsilon\beta\phi}D_{\nu\sigma\beta\phi}\Big(\bar{u}_4\Gamma^{\nu\sigma}u_2\Big).
\end{align}
Using the Dirac equation along with the structures of the propagator and vertices (and dropping terms proportional to $B(t)$), we can rewrite this as
\beq
\mathcal{A} = \frac{\lambda_{\mathcal{P}}^2A(t_1)A(t_2)\left[G_1(t_1, t_2)\eta^{\epsilon\phi} + G_2(t_1, t_2)k_2^{\epsilon}k_1^{\phi}\right]\varepsilon^{\alpha\beta\gamma\delta}k_{1\gamma}k_{2\delta}}{4(t_1 - m_g^2)(t_2 - m_g^2)}
\left[\bar{u}_3\Big(\gamma_{\alpha}P_{1\epsilon} + \gamma_{\epsilon}P_{1\alpha}\Big)u_1\right]
\left[\bar{u}_4\Big(\gamma_{\beta}P_{2\phi} + \gamma_{\phi}P_{2\beta}\Big)u_2\right],
\eeq
with
\beq
P_1 = \frac{p_1 + p_3}{2}, \hspace{.5in} P_2 = \frac{p_2 + p_4}{2}.
\eeq

\subsection{The Differential Cross Section}

In order to find the differential cross section, we now compute $\langle |\mathcal{A}|^2\rangle$, averaging over spins for the incoming protons and summing over the spins of the outgoing protons.  We find

\beq
\langle |\mathcal{A}|^2\rangle = \frac{1}{4}\sum_{\mathrm{spins}} |\mathcal{A}|^2
\eeq
$$
= \frac{16\lambda_{\mathcal{P}}^4A(t_1)^2A(t_2)^2\left[G_1(t_1, t_2)\eta^{\epsilon\phi} + G_2(t_1, t_2)k_2^{\epsilon}k_1^{\phi}\right]\left[G_1(t_1, t_2)\eta^{ef} + G_2(t_1, t_2)k_2^{e}k_{1}^{f}\right]\varepsilon^{\alpha\beta\gamma\delta}\varepsilon^{abcd}k_{1\gamma}k_{2\delta}k_{1c}k_{2d}}{(t_1 - m_g^2)^2(t_2 - m_g^2)^2}
$$
$$
\times \left(P_{1\alpha}P_{1\epsilon}P_{1a}P_{1e} - \frac{1}{16}\Big(t_1\eta_{\alpha a}P_{1\epsilon}P_{1e} + t_1\eta_{\alpha e}P_{1\epsilon}P_{1a} + t_1\eta_{\epsilon a}P_{1\alpha}P_{1e} + \left(t_1\eta_{\epsilon e} + k_{1\epsilon}k_{1e}\right)P_{1\alpha}P_{1a}\Big)\right)
$$
$$
\times \left(P_{2\beta}P_{2\phi}P_{2b}P_{2f} - \frac{1}{16}\Big(t_2\eta_{\beta b}P_{2\phi}P_{2f} + t_2\eta_{\beta f}P_{2\phi}P_{2b} + t_2\eta_{\phi b}P_{2\beta}P_{2f} + \left(t_2\eta_{\phi f} + k_{2\phi}k_{2f}\right)P_{2\beta}P_{2b}\Big)\right).
$$

The last two kinematic multiplicative terms come from expanding the traces over gamma matrices associated with the proton-proton-glueball vertices.  In each one, only the first term (proportional to $P_{1,2}^4$) contributes in the Regge limit.  This allows for great simplification, leaving us with
\beq
\langle |\mathcal{A}|^2\rangle = \frac{\lambda_{\mathcal{P}}^4A(t_1)^2A(t_2)^2s^4t_1t_2\Big(2G_1[t_1, t_2] - \mu G_2[t_1, t_2]\Big)^2\sin^2\theta_{34}}{4(t_1 - m_g^2)^2(t_2 - m_g^2)^2}.
\eeq
Combining this result with the expression for the total cross section in equation \eqref{eqn:cross} yields the differential cross section
\beq
\label{eqn:sigma}
\frac{d\sigma}{dt_1 \, dt_2 \, d\theta_{34}} = \left(\frac{\lambda_{\mathcal{P}}}{8\pi}\right)^4 \ln \left(\frac{s}{\mu}\right) \, \frac{A(t_1)^2A(t_2)^2s^2t_1t_2\Big(2G_1[t_1, t_2] - \mu G_2[t_1, t_2]\Big)^2\sin^2\theta_{34}}{(t_1 - m_g^2)^2(t_2 - m_g^2)^2}.
\eeq

Perhaps the most interesting aspect of this result is the dependence on the angle $\theta_{34}$.  The factor of $\sin^2\theta_{34}$ arises directly from the contraction with the epsilon tensor, and is therefore a simple consequence of breaking natural parity.  However, there is an additional possible dependence on $\theta_{34}$ in the factor $\mu$, which arises from the vertex structure.  At lower energies this process should be dominated by the double exchange of vector particles, where this additional structure does not occur.  Lower energy data, such as that of the WA102 collaboration at $\sqrt{s} = 29.1 \ \mathrm{GeV}$, shows no evidence of any angular dependence other than the overall $\sin^2\theta_{34}$ \cite{WA102, Kirk}, which is consistent with the idea that at these energies we expect the exchange of Reggeons to play a larger role in the process.  At higher energies, the amount that the $\sin^2\theta_{34}$ dependence is modified will be determined by the relative values of $G_1$ and $G_2$, which we will compute using the Sakai-Sugimoto model in section IV.

Finally, we need to relate this expression to the production of $\eta$ and $\eta'$ mesons.  The glueballs are flavor neutral, and thus should only couple to the flavor singlet pseudoscalar $\eta_0$.  This is a linear combination of $\eta$ and $\eta'$, using the mixing angle $\theta = -15.2^{\circ} \pm 0.5^{\circ}$:
\beq
|\eta_{0}\rangle = -\sin\theta|\eta\rangle + \cos\theta|\eta'\rangle.
\eeq
Equation \eqref{eqn:sigma} gives the result for production of $\eta_0$, and therefore the results for the production of $\eta$ and $\eta'$ should be
\beq
\label{exp:sigmaeta}
\frac{d\sigma_{\eta}}{dt_1 \, dt_2 \, d\theta_{34}} = \sin^2\theta \left(\frac{\lambda_{\mathcal{P}}}{8\pi}\right)^4 \ln \left(\frac{s}{\mu}\right) \, \frac{A(t_1)^2A(t_2)^2s^2t_1t_2\Big(2G_1[t_1, t_2] - \mu G_2[t_1, t_2]\Big)^2\sin^2\theta_{34}}{(t_1 - m_g^2)^2(t_2 - m_g^2)^2}
\eeq
and
\beq
\label{exp:sigmaetaprime}
\frac{d\sigma_{\eta'}}{dt_1 \, dt_2 \, d\theta_{34}} = \cos^2\theta \left(\frac{\lambda_{\mathcal{P}}}{8\pi}\right)^4 \ln \left(\frac{s}{\mu}\right) \, \frac{A(t_1)^2A(t_2)^2s^2t_1t_2\Big(2G_1[t_1, t_2] - \mu G_2[t_1, t_2]\Big)^2\sin^2\theta_{34}}{(t_1 - m_g^2)^2(t_2 - m_g^2)^2}.
\eeq
The overall factor of $\sin^2\theta$ or $\cos^2\theta$ is not the only way that the production of $\eta$ differs from the production of $\eta'$: the differing masses of the $\eta$ and $\eta'$ mesons also play a role, as we shall see more clearly once we have properly Reggeized the propagators, and computed the factors $G_1$ and $G_2$.

\section{Reggeizing the Propagators}

In the high energy limit we ``Reggeize'' the glueball propagators appearing in the differential cross section: we replace them with Pomeron propagators. We use a form motivated by the fact that Pomerons are dual to closed strings.  In \cite{Domokos:2009hm}, the amplitude for 4-string scattering in flat space bosonic string theory was considered, assuming that the amplitude in the curved space dual to QCD retains the general structure of the flat space amplitude, but with values for parameters such as the trajectory's slope and intercept allowed to vary.  We begin by reviewing that treatment, and then discuss the ``naive'' Reggeization of the central production process, assuming we separately Reggeize each of the propagators.  We then examine the behavior of the 5-tachyon amplitude in bosonic string theory, as developed in \cite{Herzog}.  Finally, we  propose a modification of the naive Reggeization procedure motivated by the 5-string amplitude.

This procedure relies crucially on the assumption that the curved space string scattering process involves amplitudes which essentially have the same form as flat-space amplitudes, with the entire effect of space-time curvature and extra dimensions encapsulated in the Regge trajectory parameters, which are left arbitrary. This is clearly an approximation: for instance, it has been shown that long spinning strings receive corrections to their Regge trajectories which are not linear in $J$ \cite{PZ:2003,Hellerman:2013kba}.  
However, for weak spacetime curvatures and scattering processes essentially localized at a single radial position in the holographic space, the assumption should be sufficient to create a reasonable model.

\subsection{Review of Elastic Proton-Proton Scattering}

The starting point in analyzing proton-proton scattering in \cite{Domokos:2009hm} is the crossing-symmetric Virasoro-Shapiro amplitude,
\beq
\mathcal{A} = 2\pi C \ \frac{\Gamma[-a(t)]\Gamma[-a(u)]\Gamma[-a(s)]}{\Gamma[-a(s) - a(t)]\Gamma[-a(s) - a(u)]\Gamma[-a(t) - a(s)]},
\eeq
as given in \cite{JoeI}.  This is the expression for the scattering of four closed-string tachyons, but it can be modified to account for the external particles having non-zero spin through the inclusion of a kinematic pre-factor with no poles or zeroes.  Such a modification has no effect on the procedure of Reggeization so we ignore it in what follows.

In bosonic string theory in flat space, we take $a(x) = 1 + \frac{\alpha' x}{4}$, so that the mass of the tachyon is $m_T^2 = -\frac{4}{\alpha'}$.  When we use this as an ansatz for glueball exchange in 4-proton scattering, we assume $a(x)$ is a linear function related to the glueball trajectory $\alpha_c(x) = \alpha_c(0) + \alpha_c'x$ by
\beq
\label{eqn:traj}
2 + 2a(x) = \alpha_c(x),
\eeq
so that the lowest element on the trajectory (corresponding to $a(x) = 0$) is a spin-2 glueball with mass
\beq
m_g^2 = \frac{2 - \alpha_c(0)}{\alpha_c'}.
\eeq
In replacing the dependence on $u$ with dependence on $s$ and $t$, a mass shell parameter $\chi$ is introduced so that
\beq
\label{eqn:naivechi}
a(s) + a(t) + a(u) \equiv \chi =  \frac{\alpha_c'}{2}\left[4m_p^2 - 3m_g^2\right].
\eeq
By then comparing the lowest t-channel pole with the Regge limit of this expansion, we obtain the proposed replacement in the Regge limit
\beq
\label{eqn:naive}
\frac{1}{t - m_g^2} \hspace{.2in} \Rightarrow \hspace{.2in} \frac{\alpha_c' \Gamma[-\chi]\Gamma\left[1 - \frac{\alpha_c(t)}{2}\right]}{2\Gamma\left[\frac{\alpha_c(t)}{2} - 1 - \chi\right]}  \, \left(-\frac{i\alpha_c' s}{2}\right)^{\alpha_c(t) - 2}.
\eeq
The net effect of what we have done in moving from the bosonic string theory result to the proposed Reggeization of the glueball propagator is to introduce the factor $\chi$ (which in bosonic string theory in flat space would be equal to $-1$), and replace the bosonic trajectory $a(x) = 1 + \frac{\alpha' x}{4}$ with the glueball trajectory, according to equation \eqref{eqn:traj}.  This procedure (and what follows in the analysis of the 5-string amplitude) has limitations.  However, it has the advantage of maintaining important features such as the crossing symmetry and general Regge behavior of the Virasoro-Shapiro amplitude, while allowing for some phenomenologically motivated adjustments.

\subsection{Naive Reggeization}

The naive Reggeization of the $2 \rightarrow 3$ scattering process would involve simply replacing the two glueball propagators in our amplitude according to equation \eqref{eqn:naive}.  The result would be

\beq
\frac{1}{(t_1 - m_g^2)(t_2 - m_g^2)} \hspace{.2in} \Rightarrow
\eeq
$$
 \left(\frac{\alpha_c'}{2}\right)^2\Gamma[-\chi]^2 \left(-\frac{i\alpha_c' s_1}{2}\right)^{\alpha_c(t_1) - 2}\left(-\frac{i\alpha_c' s_2}{2}\right)^{\alpha_c(t_2) - 2} \frac{\Gamma\left[1 - \frac{\alpha_c(t_1)}{2}\right]\Gamma\left[1 - \frac{\alpha_c(t_2)}{2}\right]}{\Gamma\left[\frac{\alpha_c(t_1)}{2} - 1 - \chi\right]\Gamma\left[\frac{\alpha_c(t_2)}{2} - 1 - \chi\right]}.
 $$

In order to insert this into the differential cross section, we must compute its magnitude squared.  At the same time, we will make the Regge limit approximation $s_1 \approx s_2 \approx \sqrt{s\mu}$.  This gives

\beq
\frac{1}{(t_1 - m_g^2)^2(t_2 - m_g^2)^2} \hspace{.1in} \Rightarrow
\eeq

$$
\frac{\Gamma[-\chi]^4}{s^4}\left(\frac{2}{\alpha_c' \mu}\right)^4\left(\frac{\alpha_c'^2 s\mu}{4}\right)^{\alpha_c(t_1) + \alpha_c(t_2)} \frac{\Gamma\left[1 - \frac{\alpha_c(t_1)}{2}\right]^2\Gamma\left[1 - \frac{\alpha_c(t_2)}{2}\right]^2}{\Gamma\left[\frac{\alpha_c(t_1)}{2} - 1 - \chi\right]^2\Gamma\left[\frac{\alpha_c(t_2)}{2} - 1 - \chi\right]^2}.
$$

It is interesting to note that this expression has a significant dependence on the kinematic parameter $\mu$.  This would be in addition to the dependence that arises from the structure of the coupling.  Phenomenologically, this would complicate an experimental signature associated with the angular ($\theta_{34}$) dependence of the scattering cross section.  It would also affect the ratio of the production of $\eta$ to $\eta'$ mesons (through dependence on the mass of the centrally produced meson), which would otherwise be governed almost entirely by the mixing angle.

\subsection{5-Tachyon String Amplitude in the Regge Limit}

Naive Reggeization does not take into account the fact that in the dual picture, we should be looking at a 5-string scattering amplitude.  In bosonic string theory in flat space, the five closed string tachyon amplitude can be written
\beq
\mathcal{A} = C\int d^2 u d^2 v |u|^{-2a(t_1) - 2}|v|^{-2a(t_2) - 2}|1 - u|^{-2a(s_1) - 2}|1 - uv|^{2a(s_1) + 2a(s_2) - 2a(s) - 2},
\eeq
where $a(x) = 1 + \frac{\alpha' x}{4}$ is the bosonic closed string Regge trajectory (again).  Unlike the 4-string amplitude, this cannot be computed in closed form.  However, it can be approximated in the Regge regime in two different limits:  $\frac{\alpha'\mu}{4}$  large, and  $\frac{\alpha'\mu}{4}$ small \cite{Herzog}.  In the former scenario, we obtain
\beq
\mathcal{A} \approx 4\pi^2 C \left(-\frac{i\alpha' s_1}{4}\right)^{2a(t_1)}\left(-\frac{i\alpha' s_2}{4}\right)^{2a(t_2)}\frac{\Gamma[-a(t_1)]\Gamma[-a(t_2)]}{\Gamma[a(t_1) + 1]\Gamma[a(t_2) + 1]}.
\eeq
This is essentially just the product of two separate Reggeized propagators, which suggests ``naive Reggeization.''  On the other hand, if we assume $\frac{\alpha'\mu}{4}$ is small, we obtain
\beq
\mathcal{A} \approx -4\pi^2 C \Bigg\{\left(\frac{s}{s_2}\right)^{2a(t_1)}\left(\frac{-i\alpha' s_2 }{4}\right)^{2a(t_2)} \, \frac{\Gamma[-a(t_1)]\Gamma[a(t_1) - a(t_2)]}{\Gamma[1 + a(t_1)]\Gamma[1 + a(t_2) - a(t_1)]}
\eeq
$$
\ + \ \left(\frac{s}{s_1}\right)^{2a(t_2)} \left(\frac{-i \alpha' s_1 }{4}\right)^{2a(t_1)} \, \frac{\Gamma[-a(t_2)]\Gamma[a(t_2) - a(t_1)]}{\Gamma[1 + a(t_1)]\Gamma[1 + a(t_1) - a(t_2)]} \, \Bigg\}.
$$

This expression is somewhat more complicated than the naive result.  The value of $\mu$, given in equation \eqref{eqn:eta}, is primarily determined by the mass of the centrally produced meson: $\mu \sim m_5^2$.  Based on fitting to proton-proton scattering in \cite{Domokos:2009hm}, we also know $\alpha_c' = 0.3 \ \mathrm{GeV}^{-2}$.  This gives $\frac{\alpha_c'\mu}{2} \sim 0.05$ for central production of the $\eta$ meson, and $\frac{\alpha_c'\mu}{2} \sim 0.14$ for central production of the $\eta'$ meson.  It is clearly more reasonable to use the approximation that $\frac{\alpha_c'\mu}{2}$ is small, which is not consistent with naive Reggeization.  We will therefore use the form of the 5-string amplitude as a guide to create a modified Reggeization scheme for the glueball propagators in central production processes.

\subsection{Modified Reggeization}

In analogy to the 4-string procedure, we propose a modified form of the Reggeized double glueball propagator in the
small $\frac{\alpha_c'\mu}{2}$  limit as
\beq
\frac{1}{(t_1 - m_g^2)(t_2 - m_g^2)} \hspace{.2in} \Rightarrow  \hspace{.2in} \left(\frac{\alpha_c'}{2}\right)^2\Gamma[-\chi]^2\left(-\frac{i\alpha_c' s}{2}\right)^{-2}
\eeq
$$
\times  \Bigg\{ \left(\frac{s}{s_2}\right)^{\alpha_c(t_1)}\left(\frac{-i \alpha_c' s_2 }{2}\right)^{\alpha_c(t_2)} \, \mathcal{W}(t_1, t_2)
\ + \ \left(\frac{s}{s_1}\right)^{\alpha_c(t_2)}\left(\frac{-i \alpha_c' s_1 }{2}\right)^{\alpha_c(t_1)}  \mathcal{W}(t_2, t_1)\Bigg\}.
$$
where we define
\beq
\mathcal{W}(t_1, t_2) = \frac{\Gamma\left[1 - \frac{\alpha_c(t_1)}{2}\right]\Gamma\left[\frac{\alpha_c(t_1)}{2} - \frac{\alpha_c(t_2)}{2}\right]}{\Gamma\left[\frac{\alpha_c(t_1)}{2} - 1 - \chi\right]\Gamma\left[\frac{\alpha_c(t_2)}{2} - \frac{\alpha_c(t_1)}{2} - \chi\right]}.
\eeq
Here, we maintain the same ratio of gamma functions, but we modify the trajectory to be the glueball trajectory.  This creates the correct pole structure.  We also introduce factors of $\chi$  according to their appearance in equation \eqref{eqn:naive}.  However, it is not clear what value  $\chi$ should assume.  The value found in equation \eqref{eqn:naivechi} may not be appropriate here: if $\chi$ arises from mass-shell relationships between Mandelstam variables, it should surely depend on the mass of the centrally produced meson.  In order to determine the Reggeized double propagator formally, we would need to rewrite the original 5-string amplitude in a form where the appropriate crossing symmetries are manifest, and then rewrite the expression in terms of the five independent Mandelstam variables $\{s, s_1, s_2, t_1, t_2\}$, using appropriate mass-shell relations. Nevertheless, changing the value of $\chi$ mostly rescales the Reggeized double propagator without significantly affecting its functional form, provided it is negative and of order $\mathcal{O}(1)$.  We will therefore use the value given in equation \eqref{eqn:naivechi} when simulating central production in Section V.

As before, we now compute the magnitude squared Reggeized propagators and insert $s_1 \approx s_2 \approx \sqrt{s\mu}$, as dictated by the Regge limit.  This gives

\beq
\label{eq:5stringReg}
\frac{1}{(t_1 - m_g^2)^2(t_2 - m_g^2)^2} \hspace{.2in} \Rightarrow \hspace{.2in} \frac{\Gamma[-\chi]^4}{s^4} \left(\frac{\alpha_c' s}{2}\right)^{\alpha_c(t_1) + \alpha_c(t_2)}
\eeq

$$
 \times \left\{\left[\frac{\alpha_c'\mu}{2}\right]^{\alpha_c'(t_2 - t_1)}\mathcal{W}^2(t_1, t_2) \ + \ \left[\frac{\alpha_c'\mu}{2}\right]^{\alpha_c'(t_1 - t_2)} \mathcal{W}^2(t_2, t_1) \ + \ 2\cos\left[\frac{\pi \alpha_c'(t_1 - t_2)}{2}\right]\mathcal{W}(t_1, t_2)\mathcal{W}(t_2, t_1)\right\}.
$$

Again, we will focus on the dependence on the kinematic factor $\mu$: in this expression it appears only in the form $\left[\frac{\alpha_c'\mu}{2}\right]^{\pm\alpha_c'(t_2 - t_1)}$.  The Reggeization suppresses values of $t_1$ and $t_2$ that are significantly nonzero, and so this dependence on $\mu$ is very weak, in contrast with what is implied by naive Reggeization.  Therefore, using the modified Reggeization scheme, we should expect almost all of the $\theta_{34}$ dependence to come from the structure of the glueball-glueball-pseudoscalar vertex.  Similarly, we do not expect significant differences between the production of $\eta$ versus the production of $\eta'$ to arise from the Reggeized propagators in this scheme, because there is no strong dependence on $m_5$.  Thus, using a Reggeization motivated by the 5-string amplitude is not only more consistent with string theoretic models and with the approximate value of $\alpha_c'\mu$, it also leads to cleaner and more robust predictions for the scattering behavior, making it easier to identify experimentally.

\section{Low Energy Couplings from Holographic QCD}

The general structure of central production processes can be determined from symmetry considerations and the assumption that Regge-regime scattering is well-modeled by the exchange of Regge trajectories. However, the precise structures and values of the couplings are model-dependent. Assuming that the coupling of a full Regge trajectory is completely determined by the coupling of its lightest state, we need only determine the coupling of the lightest state on each trajectory via some low-energy QCD framework -- in this case, holographic QCD.

Holographic QCD (or AdS/QCD) relies on the conjecture that there exists a gauge-string duality between QCD in 4d, and a 5d theory of strings in an aAdS spacetime. In the limit of small AdS curvature and small string coupling,
the string theory reduces to classical supergravity. This corresponds to the limit of large $\lambda=N_cg_{YM}^2$ and large $N_c$ in the gauge theory. One can therefore study strongly coupled QCD in the large $N_c$, large $\lambda$ limit using solutions to the classical supergravity equations of motion in aAdS space.

While the original AdS/CFT correspondence dealt with conformal field theories having continuous spectra, by making an appropriate choice of 5d background geometry, one can produce  a confining dual theory with a discrete spectrum.
Each supergravity field can be decomposed in a Kaluza-Klein-like tower of wavefunctions dependent on the 4d field theory's coordinates $x$, and the 5th (``holographic'') coordinate $U$: $\Phi(x,U)=\sum_n \phi_n(x)\varphi_n(U)$. Evaluating the supergravity action on these solutions, we find an effective 4d Lagrangian with an infinite number of couplings between the 4d states $\phi_n(x)$. These states correspond to towers of mesons and glueballs, each having the same quantum numbers, but different masses.

In essence, the supergravity limit gives us the first state on each of an infinite set of Regge trajectories: the lightest mass modes give the first state on primary Regge trajectories, while the more massive states in the KK tower give states on the daughter trajectories. In the supergravity limit, these are the only states we see, and states which are higher up on these Regge trajectories have infinite mass (so the Regge slopes are strictly infinite).  Of course, this limit does not accurately represent real low-energy QCD, in which no separation of scales exists between the daughter trajectories and higher spin states.

For our purposes, it will be sufficient to consider the first state on a few specific Regge trajectories regardless of the value of $\lambda$, so this shortcoming of AdS/QCD frameworks will not affect our analysis directly.  Meanwhile, models of holographic QCD have a great deal to offer.  In particular, the ``top-down'' versions of these models (like the Sakai-Sugimoto framework \cite{Sakai:2004cn} described below) have fewer free parameters than generic phenomenological frameworks.  Both bottom-up \cite{DaRold:2005zs,Erlich:2005qh} and top-down \cite{Sakai:2004cn} models  have been studied extensively in the low-energy regime, where they produce impressive
matching to experimental data (e.g. meson masses, coupling constants, etc.) and to lattice results. They also have the nice feature of incorporating older phenomenological models (such as vector meson dominance) as a natural consequence of their fifth or ``holographic'' dimension.

At best, predictions based on holographic frameworks can help evaluate the (heuristic) success of AdS/QCD models.  At worst, AdS/CFT provides phenomenological ans\"atze which may prove more successful than more commonly-used frameworks for fitting scattering data. Model-independent checks are clearly the most interesting. It is for this reason that we focus on central production of the $\eta$, whose low-energy coupling to glueballs is fixed by general anomaly-cancellation arguments in string theory (and the gravitational anomaly in QCD) and therefore depends relatively weakly on the details of the holographic model in question.

In this section we will describe the holographic QCD predictions for the proton-proton-$2^{++}$ glueball and $\eta/\eta'$-$2^{++}$ glueball couplings relevant to the central production of $\eta/\eta'$ in the Regge regime. We work exclusively in the well-studied
Sakai-Sugimoto model \cite{Sakai:2004cn}, which we now review.

\subsection{Overview of the Sakai-Sugimoto Model}

The Sakai-Sugimoto model of \cite{Sakai:2004cn} is a ``top-down'' holographic QCD framework. In contrast to the more phenomenologically-oriented ``bottom-up'' models of \cite{Erlich:2005qh, DaRold:2005zs}, the Sakai-Sugimoto model uses a $D$-brane configuration in 10d supergravity to mimic the most important features of low-energy QCD: confinement and chiral symmetry breaking. A stack of $N_c$ $D4$-branes provides the color symmetry group; stacks of parallel $N_f$ $D8$- and $\overline{D8}$-branes intersect the $D4$-branes along $3+1$ directions, generating the chiral symmetry, $U(N_f)_L\times U(N_f)_R$. In the large $N_c$ limit, we can replace the $D$4-branes with the corresponding supergravity background, which includes a Ramond-Ramond 3-form $C_3$ and a dilaton $\phi$:
\begin{align}
\label{eq:SSbackground}
ds^2=G_{MN}dx^Mdx^N=\left( \frac{U}{R}\right)^{3/2}\left( \eta_{\mu\nu}dx^\mu dx^\nu + f(U)d\tau^2 \right)+\left( \frac{U}{R}\right)^{-3/2}\left( f(U)^{-1}dU^2+U^2d\Omega_4^2\right)\\
e^{\phi}=g_s\left( \frac{U}{R}\right)^{3/4}~,\hspace{.4in} F_4=dC_3=\frac{2\pi N_c}{V_4}\epsilon_4~,\hspace{.4in} f(U)\equiv 1-\frac{U_{KK}^3}{U^3}~.
\end{align}
The coordinates $x^\mu$ denote the flat ``field theory'' directions, with $\mu=0, 1,2,3$. $R$ is the curvature, $d\Omega_4^2$ denotes the metric on an  $S^4$ transverse to the branes, $\epsilon_4$ denotes its volume form, and $V_4=8\pi^2/3$ is the volume of a unit $S^4$.  The $D4$-branes wrap an $S^1$ on finite radius (denoted by the $\tau$ direction, with $\tau\sim\tau+2\pi M_{KK}^{-1}$). To avoid a conical singularity, the radial coordinate $U$ is bounded from below as $U\ge U_{KK}$; this generates confinement in the field theory. Anti-periodic boundary conditions  on the fermionic supergravity modes, meanwhile, serve to fully break the supersymmetry of the system.

It is useful to relate the inverse radius of the $S^1$, $M_{KK}$, the asymptotic curvature $R$,  open string coupling $g_s$ and length $l_s$,  to parameters in the dual field theory:
\begin{align}
M_{KK}=\frac{3U_{KK}^{1/2}}{2R^{3/2}}\hspace{.4in} g_{YM}^2=2\pi M_{KK}g_sl_s\hspace{.4in} R^3=\pi g_sN_cl_s^3=\frac{g_{YM}^2N_cl_s^2}{2M_{KK}}~.
\end{align}
In the end, all of the physical observables for which we make predictions can be expressed in terms of $M_{KK}$ and $g_{YM}$, which
are thus the \textit{only} free parameters of the model.  We will compute  the $\rho$ meson mass and the pion decay constant in terms of the parameters $M_{KK}$ and $g_{YM}$ (detailed below), and then use experimentally observed values for
these quantities ($m_\rho=776$ MeV, $f_\pi=93$ MeV) to fix the model parameters\footnote{In practice it will prove more convenient to express all observables in terms of $M_{KK}$ and $f_\pi$ (not $g_{YM}$), but this is just a matter of trivial algebraic manipulation.}.
 As computed in \cite{Sakai:2004cn},
\begin{align}
m_\rho=0.67 M_{KK}\hspace{.4in}\text{and}\hspace{.4in} f_\pi^2=\frac{1}{54\pi^4}g_{YM}^2N_c^2M_{KK}^2~.
\end{align}
While somewhat greater accuracy could be achieved by performing a global fit to all of the low-energy QCD data at our disposal, we are in any case looking for rough estimates for the computed parameters, as the results of AdS/QCD are heuristic at best.

If we assume that $N_f\ll N_c$, we can treat the $D8$ and $\overline{D8}$ stacks as probe branes in the $D4$ background (that is, we neglect their back-reaction with the $D4$ geometry).  The full action for the system thus includes the background supergravity action $S_{grav}$ for closed string modes, the DBI action for the open string modes on the probe branes, $S_{DBI}$, and the Ramond-Ramond action $S_{RR}$ which we discuss in the next subsection.  Thus, we have
\begin{align}
\label{eq:fullSS}
S&=S_{grav}+S_{DBI}+S_{RR},\\
S_{grav}&=\frac{1}{2\kappa_{10}^2}\int d^{10}x~ \sqrt{-g}\left\{ e^{-2\phi}\left( R+4(\nabla\phi)^2\right)-\frac{(2\pi)^4l_s^2}{2\cdot 4!}F_4^2 \right\},\\
\label{eq:fullDBI}
S_{DBI}&=-T_8\int d^9x e^{-\phi} \tr\sqrt{-\det\left(\tilde{g}_{MN} + 2\pi\alpha' F_{MN} \right)}.
\end{align}
Here $\kappa_{10}$ is the 10-dimensional Newton constant, $T_8=(2\pi)^{-8}l_s^{-9}$ is the $D8$-brane tension, $\tilde{g}_{MN}$ is the pullback of the background metric onto the $D8$-branes, and $F_{MN}=\d_M A_N-\d_N A_M-i[A_M,A_N]$ denotes the field strength of $U(N_f)$-valued gauge fields living on the branes.  The traces run over $U(N_f)$ indices, with normalization $\tr (T^aT^b)=\delta^{ab}/2$.  We ignore any fluctuations along or dependence on the $S^4$ transverse to the $x^\mu$ and $U$ directions, and can thus simply integrate out the $S^4$ to yield an effectively 5d action.

In this background, the $D8$ and $\overline{D8}$-branes fill the $x^\mu$ and $S^4$ directions, while joining together to form a single, $U$-shaped stack having a non-trivial profile in the $(\tau,U)$-plane. This effectively breaks the chiral symmetry from $U(N_f)_L\times U(N_f)_R$ to $U(N_f)_V$.  One can find the profile of the $D8$-branes---and thus the induced metric on the branes---by extremizing the DBI action without gauge field fluctuations ($A_M=0$) in the $D4$-background.  The simplest possible  configuration is given by constant $\tau(U) = \frac{\pi}{3}R\sqrt{\frac{R}{U_{KK}}}$, where  the $D8$ and $\overline{D8}$ are located at antipodal points along the $x^4$ circle as $U\rightarrow\infty$.

The Sakai-Sugimoto model works in the limit of zero quark mass, so the breaking of chiral symmetry is entirely due to the $\langle \bar{q}q\rangle$ condensate. Since we work at high energies overall, this a reasonable approximation\footnote{To include an explicit quark mass (which, in the dual field theory, corresponds to explicitly deforming the field theory Lagrangian with a term of the type $m_q\bar{\psi}\psi$), one would need to turn on a non-trivial non-normalizable mode for the open string tachyon living
on the D-branes. }.

\subsection{Open and closed string spectra}

The closed string modes living in the bulk space correspond to glueball states, since they have no flavor indices. The open string modes, whose endpoints move on the stack of $D8$-branes, transform under the flavor symmetry and correspond to mesonic states in the dual field theory.  Concretely, these are modes of the $D8$-brane gauge field $A_M$ which is dual to the  QCD vector and axial vector currents. There also exist scalar excitations on the brane which come from transverse fluctuations of the brane profile in the background. These are artifacts of the model from the perspective of QCD, and we neglect them in what follows.

We now briefly review the relevant parts of the glueball and meson spectra arising in this framework, originally computed in \cite{Brower:2000rp, Constable:1999gb} and \cite{Sakai:2004cn}, respectively.

\subsubsection{Spin-2 glueballs}
The $2^{++}$ glueballs are dual to modes of the background graviton field $h_{MN}$ which transform like a symmetric traceless two-tensor in 4d: the $h_{\mu\nu}$ components. The full background field content  generates a rich spectrum of glueballs (described in detail in \cite{Brower:2000rp}), which furthermore matches lattice predictions with reasonable accuracy. For our purposes only the spin 2 mode is relevant, and we ignore the rest in what follows.

Consider the action $S_{grav}$ in equation~(\ref{eq:fullSS}), expanded in metric fluctuations around the background of equation~(\ref{eq:SSbackground}) with $g_{MN}=G_{MN}-h_{MN}$. In the gauge $h_{0\mu}=0$, and taking $h^\alpha_\alpha=0$, the graviton equation of motion is
\begin{align}
\label{heq}
-\frac{1}{2}\left( \frac{9f}{2}+3U\partial_Uf\right)h_{ij}+\left( f+U\partial_Uf\right) U \d_U h_{ij}+f U^2\d _U^2h_{ij}=-\frac{q^2 R^3}{U}h_{ij}~.
\end{align}
In this expression we have Fourier-transformed in the field theory directions as $h_{ij}(q,U)=\int d^4x e^{-iqx}h_{ij}(x,U)$.  Here $q^2$ is the 4-momentum, which has become a parameter in the solution. The boundary conditions are chosen such that solutions are smooth at $U=U_{KK}$ (i.e. $h_{ij}'(q,U_{KK})=0$), and normalizable as $U\rightarrow\infty$ ($h_{ij}(q,\infty)=0$).  Only particular values of $q^2$ satisfy \eqref{heq} with these boundary conditions, yielding a discrete spectrum of spin $2$ resonances. In terms of these resonances we can write
\begin{align}\label{eq:modeexph}
 h_{ij}(x,U)=\sum\limits_{n=1}^\infty h^{(n)}_{ij}(x)\left(\frac{U}{R}\right)^{3/2}T_n(U)~,
\end{align}
where $m_n^2$ is the mass of the $n$-th resonance $h^{(n)}_{ij}$, and the wavefunctions $T_n(U)$ satisfy
\begin{align}
\d_U\left( U^4f\d_UT_n\right) =-m_n^2R^3UT_n~.
\end{align}
The lightest of these states ($m_1^2=1.57M_{KK}^2$) is the first state on the Pomeron trajectory, while the higher states lie on daughter Regge trajectories. Inserting the mode expansion equation (\ref{eq:modeexph}) into $S_{grav}$ and integrating over $U$ yields a 4d action for the modes $h_{ij}^{(n)}$. We choose to scale $T_n$ in such a way that the kinetic terms of the 4d fields are canonically normalized. Expanding the Einstein-Hilbert term to second order in $h_{\mu\nu}$ we find
\begin{align}
S_{grav} \supset \frac{N_c^3 M_{KK}^2 g_{YM}^2}{3^5\pi^2} \sum\limits_{n=1}^\infty\int_{1}^{\infty} \frac{U}{U_{KK}} T_n^2(U)d\left(\frac{U}{U_{KK}}\right) \int d^{4}x \frac{1}{2}\eta^{\mu\nu}\eta^{\beta\delta}\eta^{\alpha\gamma}\left[\partial_{\alpha}h^{(n)}_{\delta\nu}\partial_{\mu}h^{(n)}_{\beta\gamma} - \frac{1}{2}\partial_{\alpha}h^{(n)}_{\delta\nu}\partial_{\gamma}h^{(n)}_{\beta\mu}\right] + \cdots~.
\end{align}
 We therefore require
\begin{align}
 \frac{N_c^3 M_{KK}^2 g_{YM}^2}{3^5\pi^2} \int_{1}^{\infty} \frac{U}{U_{KK}} T_m(U)T_n(U)d\frac{U}{U_{KK}}=\delta_{mn} \, .
\end{align}
As we consider only the lightest resonance here, we set $m_g^2\equiv m_1^2$, $\tT(U)\equiv T_1$, and $h_{ij}(x)\equiv h_{ij}^{(1)}(x)$ in what follows.

\subsubsection{Mesons}
Now let us consider open string excitations whose endpoints lie on the stack of $N_f$ $D8$ branes and thus transform under the flavor symmetry group. These are mesonic resonances,
parametrized in the supergravity limit by the field content of the DBI action. The meson spectrum is analyzed in great detail in  \cite{Sakai:2004cn}; we will summarize
the points relevant to our discussion here.

The gauge field $A_M$ appearing in \eqref{eq:fullDBI} is dual to the axial and vector flavor currents of QCD. Its normalizable excitations thus correspond to
the vector and axial-vector mesons ($\rho/\omega$ and $a_1/f_1$-like states), whose parity transformation
properties are dictated by whether their wavefunctions on the brane profile are even or odd under swapping the $D8$ and $\overline{D8}$ stacks, which
corresponds in QCD to swapping chirality and $U(N_f)_L\leftrightarrow U(N_f)_R$.
Even wavefunctions are dual to vectors; odd ones correspond to axial vectors.

Assuming no dependence on the $S^4$ coordinates, the DBI action equation (\ref{eq:fullDBI}) becomes
\begin{equation}
S_{D8} = -\kappa \int  d^4x d\left(\frac{U}{U_{KK}}\right) \tr\left[ \frac{3\frac{U}{U_{KK}}}{4\sqrt{\frac{U^3}{U_{KK}^3}-1}}\eta^{\mu\nu}\eta^{\rho\sigma}F_{\mu\rho}F_{\nu\sigma}
+\frac{2}{3}M_{KK}^2 \frac{U}{U_{KK}}\sqrt{\frac{U^3}{U_{KK}^3}-1}\eta^{\mu\nu}F_{\mu U}F_{\nu U}+\dots\right]~,
\end{equation}
with
\begin{equation}
\kappa = \frac{g_{YM}^2N_c^2}{108\pi^3}~.
\end{equation}
Note that there is an additional factor of $2$  in the overall coefficient (compared to the DBI action) because the $U$ coordinate only parametrizes half of the $D8$-$\overline{D8}$ profile.
In order to ensure that the mass and kinetic terms of the $4d$ action are normalizable,
we must have the field strengths $(F_{\mu Z},~F_{\mu\nu})\rightarrow 0$ as $U\rightarrow\pm\infty$, so
we seek solutions $A_M$ vanishing at large $U$. As with the graviton modes, we consider a mode expansion
\begin{align}\label{Asol}
A_\mu(x,U)&=\sum\limits_n A_\mu^{(n)}(x)\psi_n(U)~,\\
A_U(x,U)&=\sum\limits_n \phi^{(n)}(x)\varphi_n(U)~,
\end{align}
where the  $\psi_n$'s obey the equation of motion and orthogonality relation
\begin{align}
-\frac{4}{9} \frac{U_{KK}}{U}\sqrt{\frac{U^3}{U_{KK}^3}-1} \ U_{KK}d_U\left( \frac{U}{U_{KK}}\sqrt{\frac{U^3}{U_{KK}^3}-1}   \ U_{KK}d_U\psi_m \right)=-m_n^2\psi_n\quad\text{and}\quad  3\kappa\int dU
\frac{U}{ \sqrt{\frac{U^3}{U_{KK}^3}-1}  }  \psi_n\psi_m=\delta_{mn}~.
\end{align}
The massive $\varphi_n$ modes sitting in $A_U$ can be absorbed into the $A_\mu^{(n)}$ modes via a gauge transformation. There is, however,
a single massless pseudoscalar mode: the $q^2=0$ mode of the $A_U$ component of the gauge field. This mode is gauge-equivalent to the state generated by the longitudinal part of $A_\mu$, which is in turn dual to the divergence of QCD's axial flavor current. Hence the mode corresponds to the pseudoscalar mesons, which are the Nambu-Goldstone bosons associated with the broken chiral symmetry.  The states which parametrize the $SU(N_f)$ part of the flavor group are the (generalized) pions, while the $U(1)$ piece yields the $\phi_0^{U(1)}\equiv\eta_0$.
One can check that the mode
\begin{align}\label{AU}
A_U=\phi_0(x) \ 4\pi\sqrt{3N_c l_sM_{KK}}g_{YM}^2 \frac{1}{\sqrt{f}}\left(\frac{U}{U_{KK}}\right)^{-5/2}
\end{align}
satisfies the linearized equations of motion derived from the DBI action with $q^2=0$, and is furthermore orthogonal to the massive modes. The coefficient (expressed here in terms of the
field theory quantities $M_{KK}$ and $g_{YM}$)
ensures
appropriate normalization. Note that the dimensionless coordinate $U/U_{KK}\in [1,\infty)$ only parametrizes \textit{half} of the brane stack, so all integrations over $U$ on the
branes should come with an additional factor of 2.

As discussed in section II, the $\eta$ and $\eta'$ correspond to linear combinations of the $U(1)$ generator $T^0$ and the $T^8$ generator of $SU(3)$, with the $\eta'$ being mostly $T^0$, and the $\eta$ mostly $T^8$. In the above analysis, $\eta_0$ and $\eta_8$ are degenerate, since the quark masses vanish in Sakai-Sugimoto and we assume an exact $U(3)$ flavor symmetry.  This treatment neglects the additional $\eta_0$ mass generated by the anomaly in the $U(1)_A$ current in QCD.  The anomalous mass was studied in the original work of Sakai and Sugimoto \cite{Sakai:2004cn} and by \cite{Katz:2007tf} in a bottom-up QCD framework. The mass of the $\eta_0$ is non-zero when $N_c$ is large but finite, and can be derived in supergravity using the transformation properties of the background Ramond-Ramond $C_1$ potential -- which is directly analogous to the theta angle in QCD. This analysis yields \cite{Sakai:2004cn}
\begin{align}
m_{\eta_0}=\frac{1}{3\sqrt{3}\pi}\sqrt{\frac{N_f}{N_c}} M_{KK} (N_c g_{YM}^2)~.
\end{align}
In our case, this gives $m_{\eta_0} = 802$ MeV.

We adopt a practical approach and simply include the experimentally measured values for the $\eta$ and $\eta'$ masses where necessary, noting however that proper treatment of the mass in Sakai-Sugimoto could also change the mode's wavefunction on the branes, and might lead to slightly modified couplings.

\subsection{4d Couplings}
Having described the relevant parts of the spectrum, we now turn to the meson-glueball and proton-glueball couplings which determine the structure (and magnitude) of the amplitude for producing $\eta/\eta'$ in proton-proton collisions.  In top-down AdS/QCD, couplings between the open and closed string sectors arise from both the DBI action and Ramond-Ramond (RR) actions. The former generates an interaction between the protons and the spin 2 glueball, while the latter yields a natural-parity-violating coupling of $2^{++}$ glueballs to $\eta/\eta'$. As noted in the introduction, the coefficients of these terms are completely fixed on the QCD side by requiring that correlation functions of currents reproduce the gravitational anomaly.  We find the couplings between mesons and glueballs in the 4d effective theory by  evaluating the action in equation (\ref{eq:fullSS}) on shell using the mode expansions derived in the previous subsection. Each coupling constant is related to an integral over the radial coordinate $U$.

\subsubsection{$\eta_0$-glueball coupling}

Since couplings between the $\eta_0$ and two spin 2 glueballs violate natural parity, they can \textit{only} come from a Chern-Simons term that couples bulk RR forms to D-brane fields.  The Ramond-Ramond coupling for D-branes can be derived using anomaly inflow arguments \cite{Green:1996dd}. The action takes the form
\begin{align}\label{SRR}
S_{RR}=\int_{D8} C\wedge \Tr\left[\exp\left\{ \frac{F}{2\pi} \right\}\right]\sqrt{\hat{A}(\cR)}~,
\end{align}
where  integration is over the $D8$ worldvolume.
Here $C=\sum_i C_i$ is the sum of RR form fields turned on in the background. For us, $C=C_3$, the 3-form gauge potential of equation \eqref{eq:SSbackground}. $F$ is the (Hermitian) field strength of the D-brane gauge fields, and the trace is over gauge indices. $\hat{A}(\cR)$ is the ``A-roof genus,'' a sum over Pontryajin classes ($p_i$) of the gravitational curvature two-form $\cR$,
\begin{align}
\hat{A}(\cR)= 1-\frac{1}{24}p_1(\cR)+\dots=1+\frac{1}{192\pi^2}\Tr \cR\wedge \cR+\dots
\end{align}
We are suppressing $6$-form  terms and higher not relevant to the current analysis. Here the trace is over Lorentz indices of the curvature two-form, related to the Riemann tensor as $\cR^{MN} = \frac{1}{2}R_{AB}^{\phantom{AB}MN}dx^A\wedge dx^B $.

The integral in equation \eqref{SRR} picks out the $9$-form terms in the integrand:
\begin{align}\label{SRRexp}
S_{RR}&=\int_{D8} C_3\wedge\left[  \frac{1}{768\pi^3}\Tr(F)\wedge\Tr(\cR\wedge \cR ) + \frac{1}{48\pi^3} \Tr(F\wedge F\wedge F)\right]+\dots\nonumber\\
&=\int_{D8} dC_3\wedge\left[ \frac{1}{768\pi^3} \Tr(A)\wedge\Tr(\cR\wedge \cR) + \frac{1}{48\pi^3} \omega_5(A)\right]+\dots
\end{align}
where $\omega_5(A)$ is the Chern-Simons five-form, defined by $d\omega_5=\Tr F^3$. Again, we neglect  fluctuations along the $S^4$, and we can trivially integrate it out to yield a 5d action. The second term in equation \eqref{SRRexp} yields the gauge Chern-Simons term dual to the chiral anomaly of QCD, and generates natural-parity-violating couplings among mesons \cite{Sakai:2004cn,Domokos:2009cq}.

The first term in equation \eqref{SRRexp}, meanwhile, is the one of interest for modelling Pomeron exchange, as it couples glueballs to mesons.
In coordinates, the relevant 5d coupling becomes
\begin{align}
S_{RR}\supset\frac{N_c}{1536\pi^2}\int d^5x \tilde{\epsilon}^{MNPQR} \Tr(A_M)R_{NPST}R_{QR}^{\phantom{NP}TS}~,
\end{align}
where $\tilde{\epsilon}^{MNPRQ}$ refers to the Levi-Civita tensor density.  This contribution to the classical action
in supergravity generates mixed gauge-gravitional anomalies in the
full quantum theory of the dual CFT, and has been studied in the context of holographic hydrodynamics \cite{Landsteiner:2011tf}.

Since we are only interested in the $\eta_0$ coupling to gravitons,  the most convenient gauge choice is one in which the $\eta_0$ appears only
in the $U$ component of the worldvolume gauge field, $A_U$. The only term appearing in the RR-form action is then
\begin{align}\label{eq:RRwithAU}
S_{RR}\supset \frac{N_c}{1536\pi^2} \sqrt{\frac{N_f}{2}} \int d^5x \tilde{\epsilon}^{\mu\nu\rho\sigma} A^{(0)}_U R_{\mu\nu ST}R_{\rho\sigma}^{\phantom{\rho\sigma}TS}~,
\end{align}
where we have traced over the flavor indices\footnote{Flavor group generators are normalized as $\Tr T^aT^b=\delta^{ab}/2$}. To identify
the graviton-graviton-$\eta_0$ interaction we expand equation \eqref{eq:RRwithAU} to second order in the graviton perturbation, $h_{MN}$, where
$g_{MN}=\tilde{g}_{MN}-h_{MN}$. As we are only interested in the spin 2 coupling, we can neglect terms involving $h_{UU}$ and $h_{U\mu}$, focusing solely
on the terms including $h_{\mu\nu}$ and its derivatives. After some significant algebra we obtain
\begin{align}
S_{RR}\supset  & \frac{N_c}{1536\pi^2} \sqrt{\frac{N_f}{2}}\int d^5x \tilde{\epsilon}^{\mu\nu\rho\sigma U}A^0_U\left\{ \frac{9f}{2U^2}\eta^{\alpha\beta}\eta^{\gamma\delta}\d_\mu h_{\nu\alpha}\d_\sigma h_{\rho\beta}-\frac{6f}{U}\eta^{\alpha\beta}\d_U\d_\mu h_{\nu\alpha}\d_\sigma h_{\rho\beta} \right. + \\
&\qquad\qquad\qquad\qquad\qquad\qquad\qquad+\left. 2f\eta^{\alpha\beta}\d_U\d_\mu h_{\nu\alpha}\d_U\d_\sigma h_{\rho\beta}+2\left(\frac{R}{U}\right)^3 \eta^{\alpha\beta}\eta^{\gamma\delta}
\d_\gamma\d_\mu h_{\nu\alpha}\d_\sigma (\d_\delta h_{\rho\beta}-\d_\beta h_{\rho\delta}) \right\}~.\nonumber
\end{align}
Replacing the bulk fields with the lowest terms in the mode expansions computed previously, $A^0_U=\phi(U)\eta_0(x)$ and $h_{\mu\nu}(x,U)=\left(\frac{U}{R}\right)^{3/2}\tT(U) h_{\mu\nu}(x)$, we find the 4d coupling
\begin{align}
S_{\eta_0 hh}=\int d^4x \left\{ \kappa_a \epsilon^{\mu\nu\rho\sigma}\eta_0\eta^{\alpha\beta}\d_\mu h_\nu\alpha\d_\sigma h_{\rho\beta}+\kappa_b \epsilon^{\mu\nu\rho\sigma}\eta_0\eta^{\alpha\beta}\eta^{\gamma\delta}\d_\mu\d_\beta h_{\nu\gamma}\d_\rho\left( \d_\delta h_{\sigma\alpha}-\d_\alpha h_{\sigma\delta}\right)\right\} \, ,
\end{align}
where the coefficients $\kappa_a$ and $\kappa_b$ are the integrals
\begin{align}
\label{eq:ka}
\kappa_a&=  \frac{N_c}{384\pi^2} \sqrt{\frac{N_f}{2}} \int_{U_0}^\infty dU \frac{fU^3}{R^3}(\tT')^2\phi = \frac{M_{KK}^2}{4(2\pi)^5 f_\pi^3}\sqrt{\frac{N_f}{2}}\times (1.209)=0.042 \text{GeV}^{-1}\\
\label{eq:kb}
\kappa_b&=  \frac{N_c}{384\pi^2} \sqrt{\frac{N_f}{2}} \int_{U_0}^\infty dU (\tT)^2\phi=\frac{9}{16(2\pi)^5f_\pi^3}\sqrt{\frac{N_f}{2}}\times (1.043)=0.091\text{GeV}^{-1}~.
\end{align}
Recall that we have fixed $M_{KK}$ using the mass of the $\rho$, although one could obtain greater accuracy by calculating $M_{KK}$ using a more comprehensive fit of QCD observables (i.e. meson masses and couplings).

The resulting Feynman diagram coupling for graviton
polarizations $\epsilon^{(h)}_{\phi\gamma}(k_1)$ and $\epsilon^{(h)}_{\epsilon\delta}(k_2)$
 and the $\eta_0$ polarization $\epsilon^{(\eta_0)}(p_5)$
is given by
\begin{align}
2\epsilon^{\alpha\beta\gamma\delta}k_{1\mu}k_{2\sigma}\left\{ \eta^{\phi\epsilon}(\kappa_a-\kappa_b k_1\cdot k_2)+k_1^\phi k_2^\epsilon\kappa_b\right\} \, ,
\end{align}
which appears as the central vertex in figure \ref{feynman}.  Comparing this with equation \eqref{eq:vertex}, we have
\begin{align}
G_1(t_1,t_2)= 2[\kappa_a-\kappa_b (k_1\cdot k_2)] \qquad\text{and}\qquad G_1(t_1,t_2)=2\kappa_b \, .
\end{align}
With these identifications, the vertex coupling structure that appears in the differential cross sections will be
\beq
\label{eq:vform}
2G_1(t_1, t_2) - \mu G_2(t_1, t_2) = 4\left(\kappa_a + \kappa_b\sqrt{t_1 t_2}\cos\theta_{34}\right),
\eeq
which does not depend explicitly on $m_5$.  Given that the calculation of these couplings (inaccurately) assumes $m_{\eta} = m_{\eta'} = 0$ (we will simply insert the correct values for these masses later on), it is very convenient that the coupling structure we are working with ends up independent of $m_5$, so our simplified treatment should be a reasonable approximation. Notice also that the dependence of the differential cross section on $\theta_{34}$ should be sensitive to the ratio $\kappa_a/\kappa_b$.  This does not depend on $f_\pi$, only on $M_{KK}$, so errors in the determination of $f_\pi$ will affect the results only obliquely, while modifications to $M_{KK}$ are much more important.  In addition, although the values of $\kappa_a$ and $\kappa_b$ are dependent on the overlap integrals (\ref{eq:ka}) and (\ref{eq:kb}), the basic form of equation (\ref{eq:vform}) derives only from the structure of a Chern-Simons coupling, and  should be mostly independent of the details of the Sakai-Sugimoto action.

\subsubsection{Proton-glueball coupling}
Much work has been done to understand the nature of baryons in holographic QCD and in the Sakai-Sugimoto model specifically. Strictly speaking, holography operates in the $N_c\rightarrow\infty$ limit, where the baryons are infinitely massive. Baryons are also solitonic objects in the dual (gravitational) theory: they are finite-volume D-branes. In Sakai-Sugimoto, they are $D4$-branes which wrap the $S^4$ direction and are pointlike in the $U$ and $x^\mu$ directions, but  ``dissolve'' into fields living on the $D8$-brane worldvolume. In other words, baryons are charge 1 instantons of the full $5d$ DBI action of equation \eqref{eq:fullSS}. The solutions evade analytical description, and thus are often framed in terms of an expansion in $1/\lambda$ \cite{Hata:2007mb,Hashimoto:2008zw,Cherman:2011ve}; a full (numerical) solution was found only recently \cite{Bolognesi:2013nja}.  All of the above treat the baryon as an object without spin: it is only after quantization of the collective coordinate fluctuations around these solutions that they display the properties of spin 1/2 particles -- e.g. protons.
We choose instead to treat  protons using ``effective'' fermion fields $\cB$ on the curved D8-brane worldvolume with a $U$-depended effective mass \cite{Hong:2007kx,Hong:2007ay,Hong:2007dq,Park:2008sp,Domokos:2010ma}.
The resulting coupling between protons and spin 2 glueballs  was derived in \cite{Domokos:2009hm}. We very briefly review the result here.

A single graviton couples to protons as it does to the rest of the matter living on the brane: via the 5d energy-momentum tensor $T_{MN}$ \cite{Domokos:2009hm}:
\begin{align}
S_{hpp} \supset \int d^5x\sqrt{g}h_{MN}T^{MN}(x,U)~.
\end{align}
In the limit of large $\lambda$---which amounts to the assumption that the baryons `sit' at $U=U_{KK}$ with little extent to greater $U$--  \cite{Domokos:2009hm,Domokos:2010ma} showed that one can reasonably approximate the coupling of the spin $2$ portion of the graviton $h_{\mu\nu}$ to the protons via the coupling to the 4d energy-momentum tensor, with only the value of the coupling determined by 5d action wavefunctions of the fermion modes overlapping with the graviton:
\begin{align}
S_{hpp}\propto \lambda_{\mathcal{P}}\int d^4x h_{\mu\nu} T_p^{\mu\nu} \, ,
\end{align}
where $T_p$ is the 4d stress tensor of the proton modes. In the treatment of protons as effective fermion fields, the coupling $\lambda_{\mathcal{P}}$ was
found to be $\lambda_{\mathcal{P}}=6.38~\text{GeV}^{-1}$ \cite{Domokos:2010ma}, which is the value we use below. Note, however, that the fit of a Regge regime ansatz
for proton-proton scattering, performed in \cite{Domokos:2009hm}, yielded a value of approximately $\lambda_P=8.5~\text{GeV}^{-1}$. Here we adopt values calculated in the Sakai-Sugimoto
model (rather than fits) wherever possible. The value of $\lambda_P$ only affects the magnitude of the total cross-section, which we cannot predict reliably anyway, as described below.

As in \cite{Domokos:2009hm}, we can model the behavior of the energy momentum tensor itself by considering its matrix element between proton states,
\begin{align}
\langle p',s' | T_{\mu\nu} |p,s\rangle =\bar{u}(p',s')\left[\frac{A(t)}{2}\left(\gamma^{\mu}P^{\rho} + \gamma^{\rho}P^{\mu}\right) + \frac{B(t)}{8m_p}\Big(P^{\mu}[\gamma^{\rho}, \gamma^{\nu}] + P^{\rho}[\gamma^{\mu},\gamma^{\nu}]\Big)k_{\nu} - \frac{C(t)}{m_p}\left(\eta^{\mu\rho}t + k^{\mu}k^{\rho}\right)\right] u(p,s)~,
\end{align}
where $P=(p+p')/2$ and $k=p-p'$, and $t=-k^2$. Assuming that the Sakai-Sugimoto baryon roughly takes the form of a 4d Skyrmion in the spherically symmetric hedgehog configuration, one can explicitly  compute these form factors in the large $N_c$ limit \cite{Cebulla:2007ei}. It was shown in \cite{Domokos:2009hm} that  $B(t)$ is small and slowly-varying for small $|t|$, and as noted in section II, dependence on $C(t)$ disappears in our amplitude. We can therefore neglect $B(t)$ and $C(t)$. Meanwhile, for $|t|<0.8$ GeV, $A(t)$ is well-approximated by a dipole form with $M_d=1.14$ GeV  \cite{Domokos:2009hm}.

\section{Simulating Production}

We are now ready to write down our final result for the Reggeized differential cross section and use it to simulate central production. We focus on the angular dependence of the differential cross section Reggeized according to the 5-tachyon string amplitude.  To better understand the effects of the modified Reggeization procedure, we compare this to the results using naive Reggeization as well as no Reggeization.  We are primarily concerned with $\eta$ production as our treatment of the $\eta'$ meson does not account for the instanton effects responsible for the mass splitting between $\eta$ and $\eta'$.  Additionally, in the 5-string Reggeization, the approximation that $\alpha_c'\mu$ is small is stronger for $\eta$ than $\eta'$.  Nevertheless, we  compare $\eta$ production to that of $\eta'$ to highlight important differences.  Finally, we  compare the total cross sections for $\eta$ and $\eta'$ with each type of Reggeization.

\subsection{Simulating $\eta$ Production with 5-string Reggeization}

\begin{table}
\begin{tabular}{||c|c|l||}
\hline
\hline
$\lambda_\mathcal{P}$ & $6.38 \ \mathrm{GeV}^{-1}$ & value determined using effective fermion fields  in the \\
& & Sakai-Sugimoto model, in \cite{Domokos:2010ma} \\
\hline
$\alpha_c'$ & $0.290 \ \mathrm{GeV}^{-2}$ & fit value using proton-proton scattering data \\
& & modeled with string-theory inspired amplitude in \cite{Domokos:2009hm} \\
\hline
$m_g$ & $1.485 \ \mathrm{GeV}$ & value determined using the Sakai-Sugimoto dual model in \cite{Brower:2000rp} \\
\hline
$\kappa_a$ & $0.0423 \ \mathrm{GeV}^{-1}$ & value determined using the Sakai-Sugimoto dual model in section IV  \\
\hline
$\kappa_b$ & $0.0921 \ \mathrm{GeV}^{-3}$ & value determined using the Sakai-Sugimoto dual model in section IV \\
\hline
$M_d$ & $1.17 \ \mathrm{GeV}$ & value determined in the Sakai-Sugimoto dual picture \\
& & modeling the proton as a 4-dimensional skyrmion, in \cite{Domokos:2009hm} \\
\hline
$m_p$ & $0.938 \ \mathrm{GeV}$ & known experimental value \\
\hline
$m_{\eta}$ & $0.548 \ \mathrm{GeV}$ & known experimental value \\
\hline
$m_{\eta'}$ & $0.958 \ \mathrm{GeV}$ & known experimental value \\
\hline
$\theta$ & $15.2^{\circ}$ & known experimental value \\
\hline
\hline
\end{tabular}
\caption{\label{valuestable} A table of parameter values used in the simulations.}
\end{table}

After Reggeization based on the 5-tachyon string amplitude, as given in equation (\ref{eq:5stringReg}), the tree-level differential cross section for $\eta$ reads
\begin{align}
	\frac{d\sigma}{dt_1 dt_2 d\theta_{34}} = &\sin^2\theta\of{\frac{\lambda_{\mathcal{P}}\Gamma[-\chi]}{4\pi}}^4\ln(s/\mu)\of{\frac{\alpha_c's}{2}}^{\alpha_c(t_1)+\alpha_c(t_2)}A(t_1)^2A(t_2)^2\frac{t_1 t_2 \sin^2\theta_{34}}{s^2}(\kappa_a+\kappa_b\sqrt{t_1t_2}\cos\theta_{34})^2
	\RET
	&\times
	\Bigg\{
		\sqof{\frac{\alpha_c' \mu}{2}}^{\alpha_c'(t_2-t_1)}\mathcal{W}^2(t_1,t_2)
		+
		\sqof{\frac{\alpha_c' \mu}{2}}^{\alpha_c'(t_1-t_2)}\mathcal{W}^2(t_2,t_1)
		+
		2\cos\sqof{\frac{\pi \alpha_c'(t_1-t_2)}{2}}\mathcal{W}(t_1,t_2)\mathcal{W}(t_2,t_1)
	\Bigg\}.
\end{align}
The form factor $A(t)$ is assumed to be
\beq
A(t) = \left(1 - \frac{t}{M_d^2}\right)^{-2},
\eeq
and we use
\beq
\chi = \frac{\alpha_c'}{2}[4m_p^2 - 3m_g^2], \hspace{.5in} \alpha_c(x) = \alpha_c(0) + \alpha_c'x, \hspace{.5in} \alpha_c(0) = 2 - \alpha_c'm_g^2.
\eeq
The values of the parameters arising in the equations above are given in table \ref{valuestable}.  The masses of the external particles (the protons and the pseudoscalars) and the mixing angle between $\eta$ and $\eta'$ are fixed to be their experimentally known values.  Most of the other parameters are determined using the Sakai-Sugimoto model as our low energy dual supergravity theory.  The exception is the Regge slope $\alpha_c'$, which is fit to existing proton-proton scattering data, because it cannot be extracted from the low energy supergravity theory (in which the slope is, strictly speaking, infinite).

\begin{figure}[htb!]
\begin{center}
\includegraphics[scale=.7]{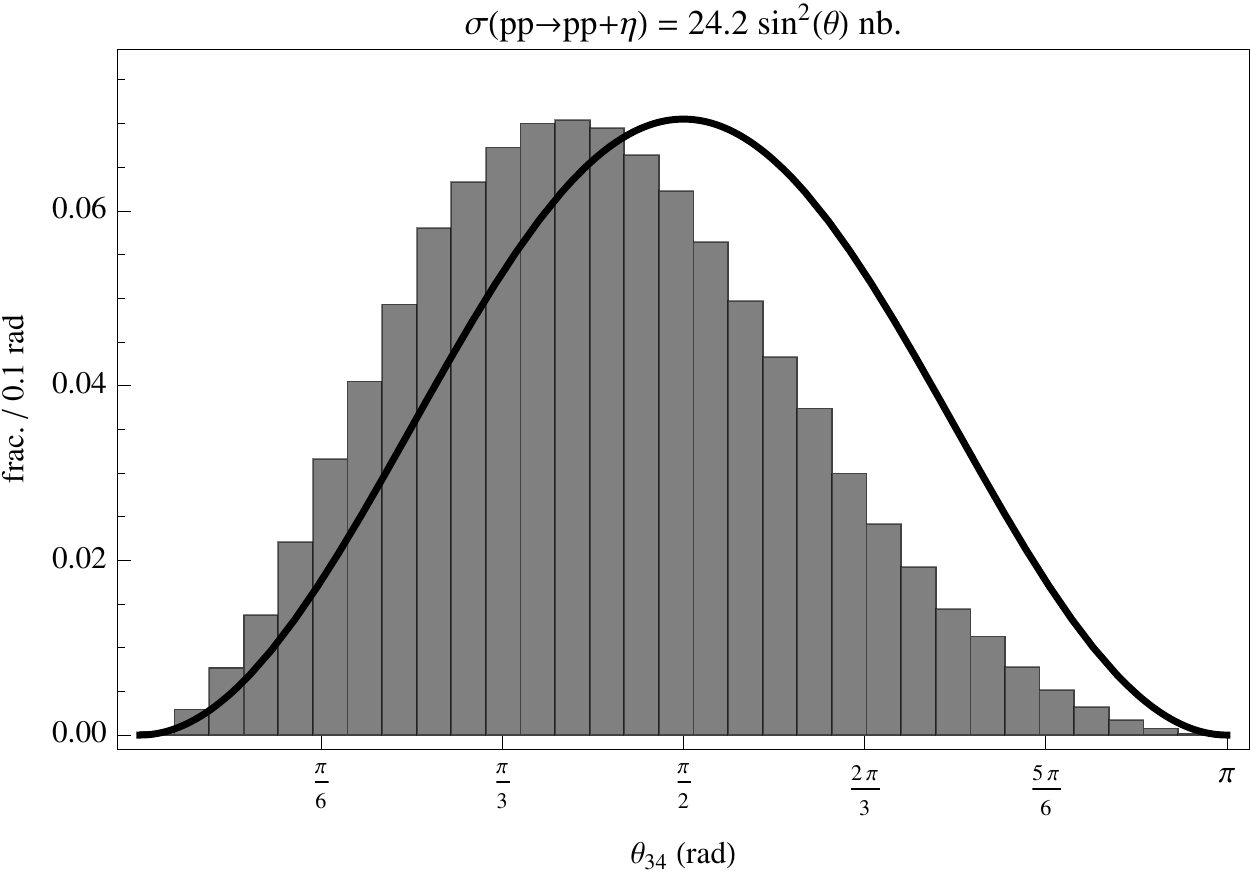}
\includegraphics[scale=.7]{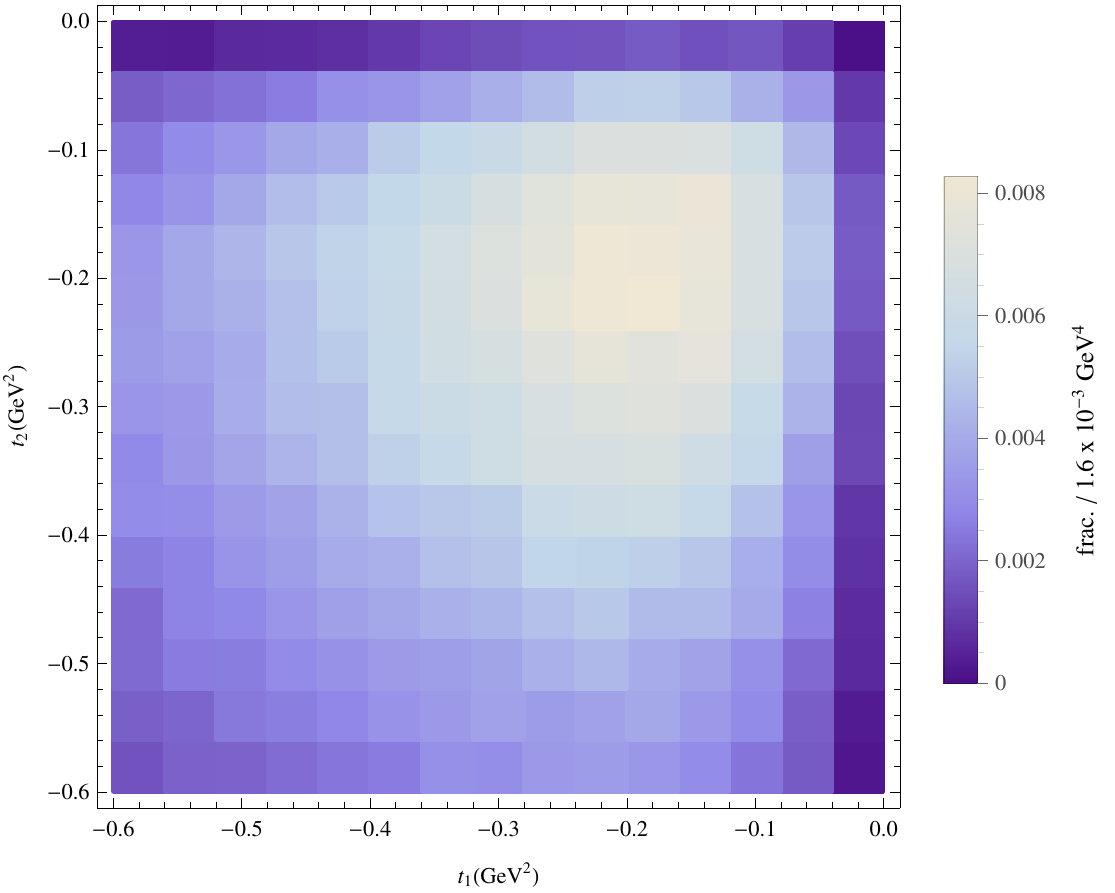}
\caption{The differential cross sections $\frac{d\sigma}{d\theta_{34}}$ and $\frac{d\sigma}{dt_1 dt_2}$, Reggeized according to the 5-tachyon string amplitude, are shown. In the angular dependence, qualitative deviations from the pure $\sin^2\theta_{34}$ behavior are visible, with a maximum at approximately 1.3 radians. In the $t_1$ and $t_2$, dependence, we see, by comparison to the un-Reggeized differential cross sections shown in figure \ref{fig:eta_unreg}, that the Reggeized double glueball propagator selects most strongly for events where $t_1$ and $t_2$ are between approximately $-0.04$ and $-0.24\,\text{GeV}^2$. Additionally, comparing the $t_{1}$ and $t_2$ dependence to the naively Reggeized results, shown in figure \ref{fig:eta_naive}, we see that the Reggeized double glueball propagator suppresses events with larger $|t_{1,2}|$ more slowly.}
\label{fig:eta_full}
\end{center}
\end{figure}

Having fixed all of the form factors, coupling constants, and masses, we proceed to simulate the differential cross section using the rejection method, at a center-of-mass (CM) energy of the WA102 data for $\eta$ central production \cite{WA102}, $\sqrt{s} = 29.1$ GeV. For $t_{1,2} < -0.6\, \text{GeV}^2$ we assume that perturbative QCD effects dominate, so we take $t_1$ and $t_2$ to range from $0.0$ to $-0.6\, \text{GeV}^2$.  At this energy the data for $\frac{d\sigma}{d\theta_{34}}$ takes a characteristic $\sin^2\theta_{34}$ shape, as a direct result of natural parity violation. However, as shown in figure \ref{fig:eta_full}, the double Pomeron exchange contribution to $\frac{d\sigma}{d\theta_{34}}$ exhibits a different, modulated $\sin^2\theta_{34}$ form due to the additional $\cos\theta_{34}$ dependence, and has a maximum visibly below $\frac{\pi}{2}$, at approximately 1.30 radians.  At this CM energy, the WA102 experiments \cite{WA102} seem to indicate that double Reggeon exchange dominates the process, as the total exclusive cross section for pions (which cannot be produced by double Pomeron exchange) is about ten times larger than the total cross section for $\eta$.  At higher energies, however, we expect that double Pomeron exchange  will dominate and that the differential cross section will exhibit modifications from the pure $\sin^2\theta_{34}$ behavior.

Let us examine the source of this deviation from $\sin^2\theta_{34}$ behavior. Besides the overall factor of $\sin^2\theta_{34}$, there are three sources of angular dependence in the differential cross section, all of which arise from the structure $\mu \approx m_5^2 - t_1 - t_2 + 2\sqrt{t_1t_2}\cos\theta_{34}$. First, we have the factor of $\ln(s/\mu)$, which comes from the kinematics and phase space of $2\to 3$ scattering. Because $\mu$ is always close to $m_5^2$ and a logarithm is a slowly varying function, the quantity $\ln(s/\mu)$ contributes to the differential cross section primarily as an overall scale factor. Second, we have the factors of $(\alpha_c'\mu)^{\pm\alpha_c'(t_2-t_1)}$ from the Reggeized double glueball propagator. However, since $|\alpha_c'(t_2-t_1)| \leq 0.174$, the quantity $(\alpha_c'\mu)^{\pm\alpha_c'(t_2-t_1)}$ will also varies quite slowly: at most 10\% over the domain of the differential cross section. Finally, we have the expression $\of{\kappa_a+\kappa_b\sqrt{t_1t_2}\cos\theta_{34}}^2$, which comes from the structure of the spin 2-spin 2-pseudoscalar vertex. This is the dominant factor modifying the $\sin^2\theta_{34}$ dependence. The ratio $\kappa_a/\kappa_b$ controls the degree of this modification, suggesting that the deviation of $\frac{d\sigma}{d\theta_{34}}$ from $\sin^2\theta_{34}$ could be used (at higher energies) as an experimental test both of the even-spin nature of the Pomeron trajectory (which is what allows such a vertex structure to exist in the first place), as well as the specific value of $\kappa_a/\kappa_b$.

\subsection{Comparison to Naive Reggeization and No Reggeization}
\begin{figure}[htb!]
\begin{center}
\includegraphics[scale=.7]{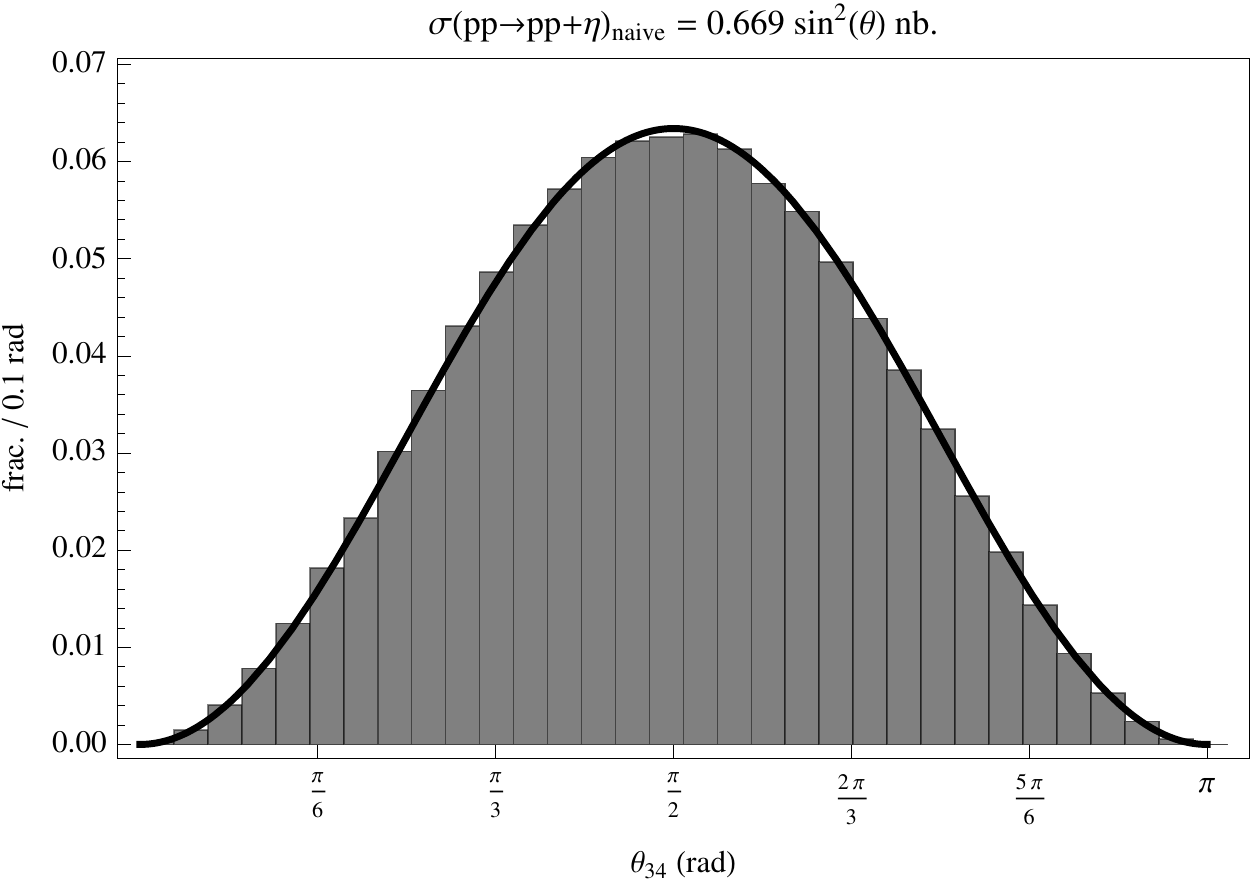}
\includegraphics[scale=.7]{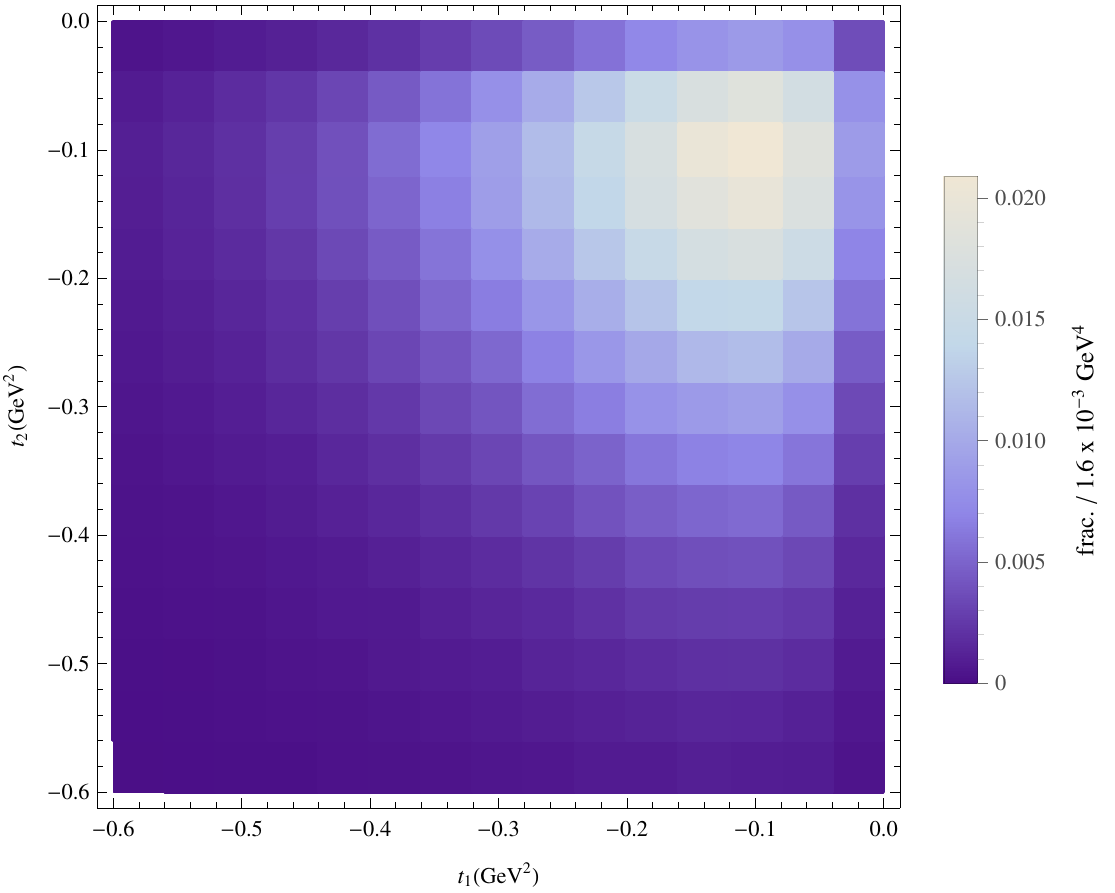}
\caption{The differential cross sections $\frac{d\sigma_{\text{naive}}}{d\theta_{34}}$ and $\frac{d\sigma_{\text{naive}}}{dt_1 dt_2}$, Reggeized naively, are shown. In the angular dependence, qualitative deviations from the pure $\sin^2\theta_{34}$ behavior of Reggeon exchange are barely visible, with a maximum at or just above $\pi/2$. In the $t_1$ and $t_2$, dependence, we see that the naively Reggeized propagators selects most strongly for events where $t_1$ and $t_2$ are  approximately $-0.1\,\text{GeV}^2$.}
\label{fig:eta_naive}
\end{center}
\end{figure}

\begin{figure}[htb!]
\begin{center}
\includegraphics[scale=.7]{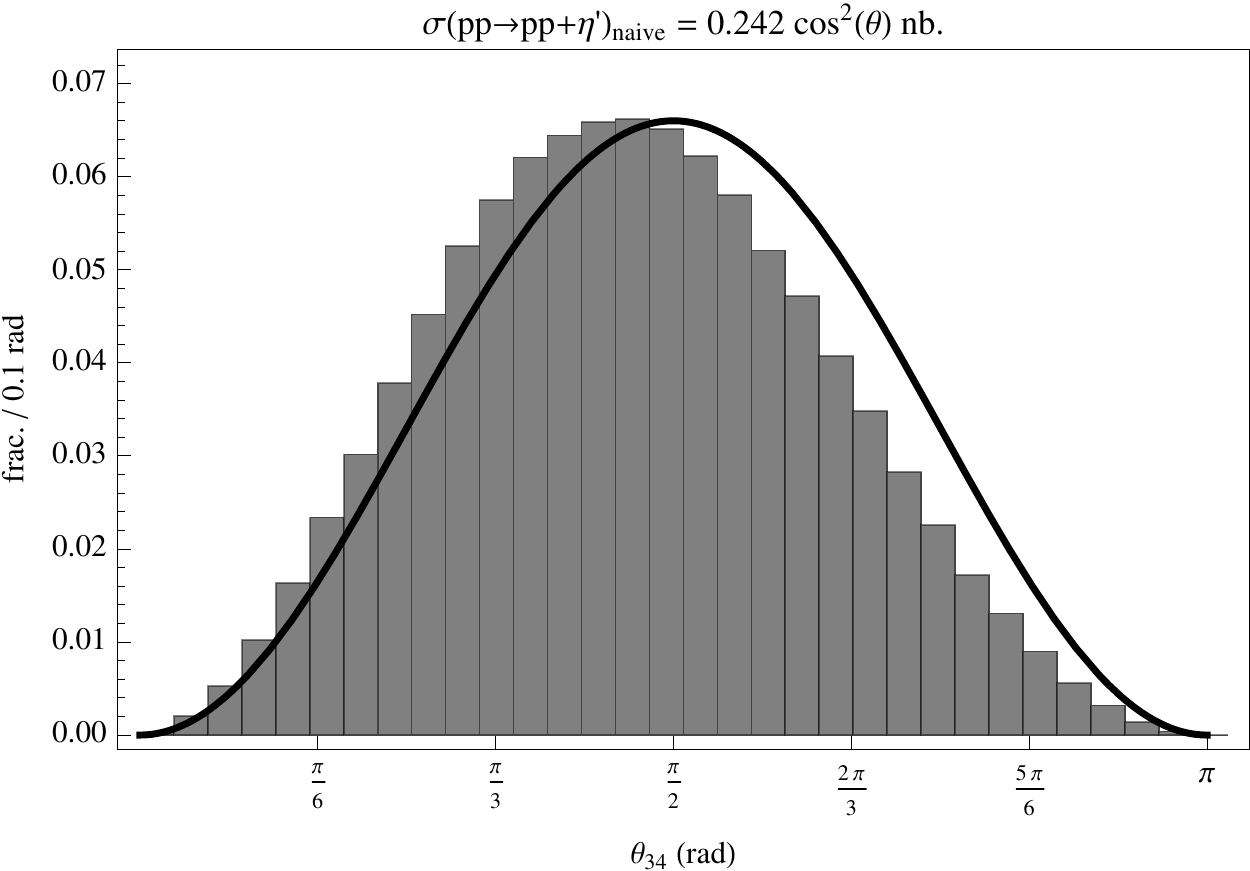}
\includegraphics[scale=.7]{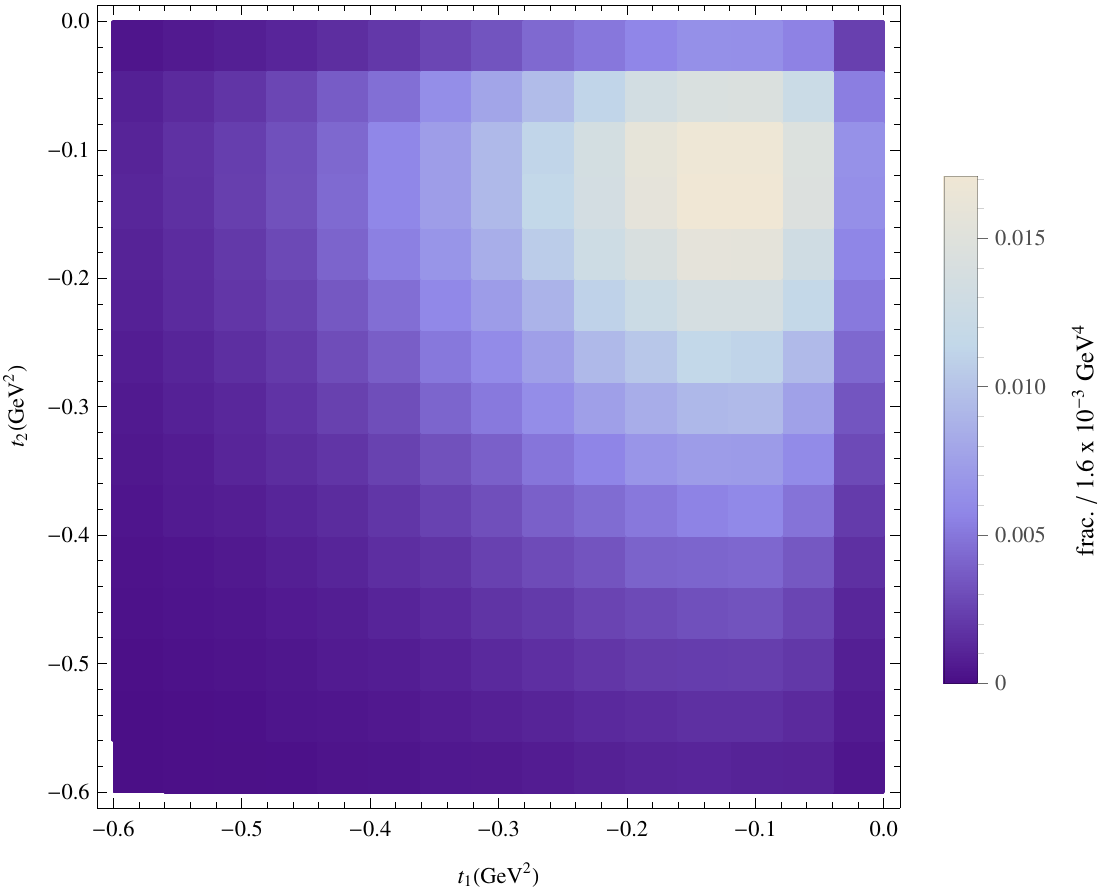}
\caption{The naively Reggeized differential cross sections for $\eta'$ production, $\frac{d\sigma_{\text{naive}}}{d\theta_{34}}$ and $\frac{d\sigma_{\text{naive}}}{dt_1 dt_2}$, are shown. With $m_5 = m_{\eta'}$, less cancelation occurs between the $\theta_{34}$ dependence from the naively Reggeized propagators and the spin 2-spin 2-pseudoscalar vertex, and the maximum is shifted below $\pi/2$.}
\label{fig:etaprime_naive}
\end{center}
\end{figure}

As discussed previously, the Reggeization scheme based on a 5-string amplitude structure has significantly different implications for the scattering process than a naive Reggeization scheme.  In order to highlight this, we can also look at simulations based on the naive Reggeization scheme, and for the un-Reggeized differential cross section.  The naively Reggeized differential cross section for $\eta$ production reads
\begin{align}
	\frac{d\sigma_{\text{naive}}}{dt_1 dt_2 d\theta_{34}} = &\sin^2\theta\of{\frac{\lambda_{\mathcal{P}}\Gamma[-\chi]}{4\pi}}^4\ln\of{s/\mu}\of{\frac{2}{\alpha_c'\mu}}^4\of{\frac{\alpha_c'^2s \mu}{4}}^{\alpha_c(t_1)+\alpha_c(t_2)} \RET
	&\times\of{\frac{A(t_1)\Gamma\sqof{1-\frac{\alpha_c(t_1)}{2}}}{\Gamma\sqof{\frac{\alpha_c(t_1)}{2}-1-\chi}}}^2
	\of{\frac{A(t_2)\Gamma\sqof{1-\frac{\alpha_c(t_2)}{2}}}{\Gamma\sqof{\frac{\alpha_c(t_2)}{2}-1-\chi}}}^2
	\frac{t_1 t_2 \sin^2\theta_{34}}{s^2}\of{\kappa_a+\kappa_b\sqrt{t_1t_2}\cos\theta_{34}}^2,
\end{align}
and the simulated angular dependence as well as $t_1$ and $t_2$ dependence are shown in figure \ref{fig:eta_naive}.  We notice immediately that $\frac{d\sigma_{\text{naive}}}{d\theta_{34}}$ shows only a very weak deviation from the $\sin^2\theta_{34}$ shape, with a maximum just greater than $\pi/2$. This effect is largely due to a partial cancellation between the angular dependence of the central vertex structure in equation (\ref{eq:vform}) and the factor $\of{\alpha_c'\mu/4}^{\alpha_c(t_1)+\alpha_c(t_2)-4}$, in the naively Reggeized propagators. This cancellation depends on the precise value of $m_5 = m_{\eta}$. If we instead look at $\eta'$ production, shown in figure \ref{fig:etaprime_naive}, this cancellation effect is less exact due to the increased value of $m_5$, and the maximum shifts to below $\pi/2$.  Additionally, the smaller deviation from $\sin^2\theta_{34}$ agrees with the $t_1$ and $t_2$ dependence produced by the naive Reggeization, which more rapidly suppresses the differential cross section for larger $|t_1|$ and $|t_2|$.

\begin{figure}[htb!]
\begin{center}
\includegraphics[scale=.7]{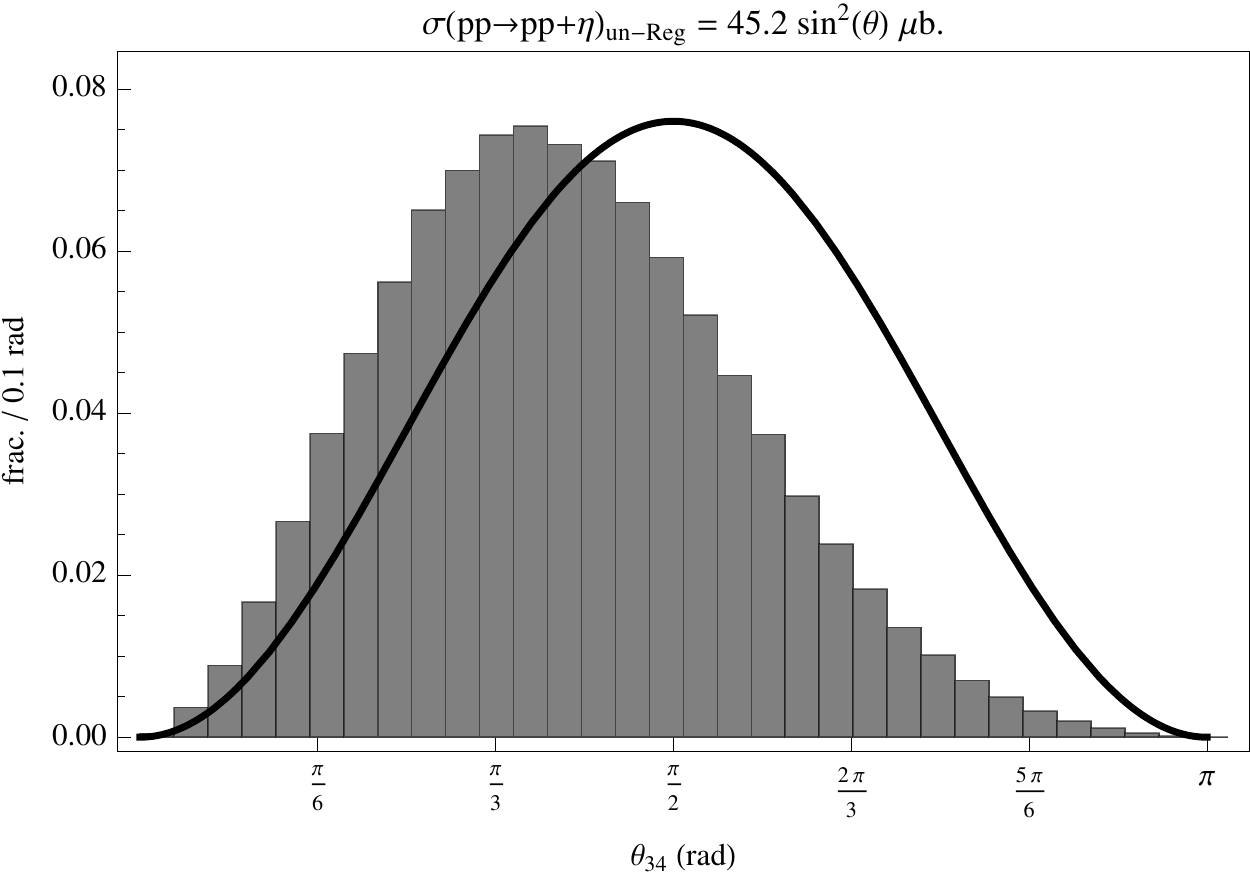}
\includegraphics[scale=.7]{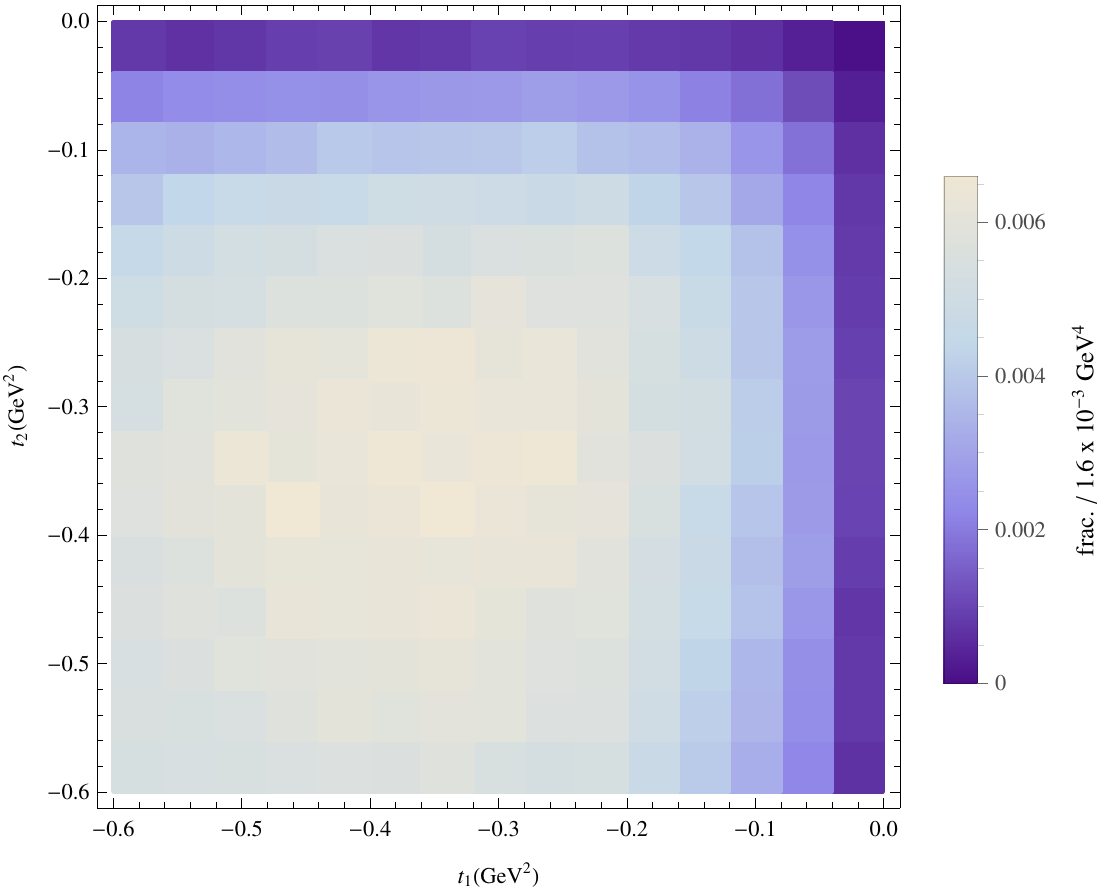}
\caption{The un-Reggeized differential cross sections $\frac{d\sigma_{\text{un-Reg}}}{d\theta_{34}}$ and $\frac{d\sigma_{\text{un-Reg}}}{dt_1 dt_2}$ are shown. The angular dependence, as compared to either of the Reggeized differential cross sections, differs more strongly from pure Reggeon exchange and takes a maximum at approximately 1.1 radians. Examining the $t_1$ and $t_2$ dependence, we see that this occurs because without Reggeization the events where larger $|t_1|$ and $|t_2|$ have more likelihood, thereby enhancing the $\theta_{34}$ dependence.}
\label{fig:eta_unreg}
\end{center}
\end{figure}

The simulation of the un-Reggeized differential cross section is shown in figure~\ref{fig:eta_unreg}. The un-Reggeized results show the strongest deviation from $\sin^2\theta_{34}$, and examining the $t_1$ and $t_2$ dependence we see that this occurs primarily because the differential cross section is largest when $|t_1|$ and $|t_2|$ are large, as opposed to being quickly suppressed in this region.  Any form of Reggeization will suppress larger $|t_1|$ and $|t_2|$, which will decrease the amount of deviation.

\subsection{Total Cross Sections for $\eta$ and $\eta'$ Production}

We can also use the simulations to compute the total cross sections for the production of $\eta$ and $\eta'$, which we can compare with each other as well as with the data from the WA102 experiment.
The tree-level total cross sections at 29.1 GeV for $\eta$ and $\eta'$ predicted by 5-string Reggeization are
\begin{align}
	\sigma(pp\to pp+\eta) = 24.2 \sin^2\theta\,\text{nb} \qquad\text{and}\qquad
	\sigma(pp\to pp+\eta') = 19.1 \cos^2\theta\,\text{nb}.
\end{align}
Thus, using 5-string Reggeization, one expects the ratio of the $\eta$ and $\eta'$ total cross sections to fall within 25\% of $\tan^2\theta$, the relationship based solely on the mixing angle.  On the other hand, the ratio of the naively Reggeized $\eta$ and $\eta'$ total cross sections is approximately $3\tan^2\theta$, due to the additional dependence on $m_5$ in this calculation.

It is also interesting to compare these cross sections to the experimental cross sections from the WA102 experiment \cite{WA102}, which found
\begin{align}
	\sigma(pp\to pp+\eta)_{\mathrm{exp}} = (3859 \pm 368) \, \text{nb} \qquad\text{and}\qquad
	\sigma(pp\to pp+\eta')_{\mathrm{exp}} = (1717 \pm 184) \,\text{nb},
\end{align}
at $\sqrt{s}=29.1$ GeV.
It would thus appear that double Pomeron exchange accounts for only about $1\%$ of the production of $\eta'$, and less than $0.1\%$ of the production of $\eta$.  However, it is worth noting that the overall cross sections calculated are sensitive to a number of factors that we have  estimated only very roughly. In particular, varying the mass-shell parameter $\chi$ by $\pm 50\%$ changes the cross sections by an order of magnitude.  However, this sensitivity is not carried over into either the dependence on $\theta_{34}$ or the ratio of the $\eta$ and $\eta'$ total cross sections. Furthermore, such a strong hierarchy is not surprising from the holographic perspective: the coefficient of the 5d gravitational Chern-Simons term that gives rise the Pomeron-Pomeron-$\eta_0$ interaction is determined by the anomaly to be $N_c/1536\pi^2$, while the pure gauge Chern-Simons term that generates the Reggeon-Reggeon-$\eta$ vertex has a coefficient of $N_c/24\pi^2$. Without
taking into account the reduction to a 4d effective theory, whose couplings are determined in part by these coefficients, in part by wavefunction overlaps (as described in the previous section), the hierarchy of Chern-Simons coefficients implies that the cross sections differ by a factor of roughly 2500! The dependence on $s$, meanwhile, favors Pomeron exchange as $s$ grows larger,
since Reggeized Pomeron propagators scale (roughly) with $s^{0.04}$ while Reggeons scale with $s^{-0.19}$\footnote{As determined from the proton-proton total cross section.}, leading to naively Reggeized central production cross sections that approximately go like $(s_1s_2)^{0.08}$ and $(s_1s_2)^{-0.38}$, respectively. It is thus clear that while understanding Reggeon exchange is essential at presently-measured energies, at high enough energies Pomeron exchange should play the dominant role.

\section{Conclusions and future directions}

We have used the ideas of holographic QCD and Regge theory to construct a model for the central production of a pseudoscalar meson via double Pomeron exchange in proton-proton collisions.   Our starting point was the tree-level process involving the t-channel exchange of massive spin-2 particles, where the forms of the vertices and propagators are dictated by Lorentz, parity and charge conjugation symmetry.  The central vertex, in particular, must violate natural parity, which leads to an overall factor of $\sin^2\theta_{34}$ in the cross section, where $\theta_{34}$ is the angle between the emerging photons in the plane transverse to the scattering process.  Assuming that the exchanged particles are spin 2 leads to additional possible structures (and additional dependence on $\theta_{34}$), that  are not present if for instance the exchanged particle is spin 1.

Motivated by gauge-string duality, we considered this process to be the low-energy limit of a 5-string scattering process, and sought to find an appropriate ``Reggeization.''  It was previously shown that in elastic proton-proton scattering, the Regge limit of the string amplitude is reasonably approximated by simply replacing low-energy propagators with a Reggeized propagator that encompasses the entire exchanged Regge trajectory, in this case a Pomeron.  We found that a naive approach based on this idea, separately replacing each propagator with a Pomeron, is not equivalent to what follows from analyzing a 5-string amplitude in the Regge limit, unless the mass of the centrally produced meson is sufficiently large. This should not be the case for the $\eta$ or $\eta'$ mesons.  We therefore proposed a modified Reggeization procedure for this scattering process.  This form of Reggeization, in particular, does not introduce significant additional dependence on the angle $\theta_{34}$ into the scattering cross section (while the naive Reggeization process would).

We then computed the low energy coupling constants using the Sakai-Sugimoto model as the supergravity limit of the dual theory.  Here, the natural-parity-violating central vertex arises from a Chern-Simons action which reproduces the gravitational anomaly in QCD.  The values of the two coupling constants involved were computed as overlap integrals depending on the modes of the graviton in the bulk and the vector field on the D8-brane.  The Chern-Simons action leads to a vertex structure that includes more information than can be inferred using symmetry arguments alone.  This information is likely to be relatively model independent; any QCD dual theory must contain parity violating Chern-Simons terms that reflect the chiral gravitational anomaly of QCD. These five-dimensional Chern-Simons couplings are universal, although some weak model dependence enters through the wave functions of the glueballs and $\eta$ and $\eta'$ mesons.

Finally, we generated simulations of the scattering process at the energy $\sqrt{s} = 29.1$ GeV, using our Reggeization procedure for the propagators and the values of the coupling constants derived from the Sakai-Sugimoto model.  We saw a clear shift of the differential cross section $\frac{d\sigma}{d\theta_{34}}$ away from a pure $\sin^2\theta_{34}$ profile.  Experimental data at this energy shows no such deviation, supporting the idea that at this energy double Reggeon exchange dominates the process.  We also computed the total cross sections for both $\eta$ and $\eta'$ production.  Using our Reggeization procedure, we found that the ratio is primarily determined by the mixing angle between $\eta$ and $\eta'$: the glueballs couple only to the flavor singlet.

Given that the existing central production data (e.g. from the WA102 \cite{WA102} experiment) lies squarely in the regime where Reggeon exchange seems to dominate, a crucial next step in this analysis is to create a model that  also incorporates double Reggeon exchange.  It will be an important zeroth-order check of our methods to see whether holographic calculations which include Reggeons give the right ballpark estimate for the total cross section. Since double Reggeon process should dominate at this energy,  the dependence of the differential cross section on $\theta_{34}$ should be a simple $\sin^2\theta_{34}$ profile\footnote{Preliminary analysis of the 5-open-string amplitude indicates that significant modifications to the $\theta_{34}$ dependence from Reggeization are unlikely; the amplitude in the small $\mu$/Regge limit has only weak dependence on $\mu$, just as for the 5-closed-string amplitude.}.
It would be interesting to analyze the behavior of the full model at increasing center-of-mass energy so as to pinpoint where significant deviations from $\sin^2\theta_{34}$ begin to arise.

The present analysis could also be made more consistent.  The Reggeization of propagators was somewhat ad hoc: we used the 5-string scattering amplitude for flat space bosonic strings as a starting point, but did not take into account the modifications of the mass-shell conditions in a well-motivated way.  It would be interesting to take a more systematic approach, particularly as this might lead to additional dependence on the mass of the centrally produced meson.  We could also use a more accurate treatment of the proton in the dual model that better accounts for the 5d structure of the process instead of simply relying on the  Skyrmion solution for both the form factor and the coupling constant between the protons and the glueballs.  However, the recent numerical work of \cite{Bolognesi:2013nja} on exact skyrmion solutions suggests that this may not be a reasonable approach.  A numerical analysis to determine these factors would be more appropriate, and might yield further insights.

Overall, our results suggest that  the central production of pseudoscalar mesons in very high energy proton-proton collisions could provide interesting insights into the success of string/gravity duals for QCD. Though the details of the production rate are model-dependent, the central ingredient -- a natural parity-violating coupling between glueballs and pseudoscalar mesons required by the gravitational contribution to the chiral anomaly -- is not.

\begin{acknowledgements}
SKD is supported by the NYU Postdoctoral and Transition Program for Academic Diversity.
JH acknowledges the support of NSF grant 1214409.  NM and NA acknowledge the support of the Reed College Summer Scholarship Fund.
\end{acknowledgements}

\begin{appendix}

\section{Regge Limit and Phase Space}

\subsection{The Phase Space}

Calling $\langle |\mathcal{A}|^2\rangle$ the spin-averaged amplitude squared, we know the total cross section should be
\beq
\sigma = \frac{1}{64\sqrt{(p_1\cdot p_2)^2 - m_p^4}}\int \frac{\langle|\mathcal{A}|^2\rangle (2\pi)\delta\left(2E - \sqrt{{\bf p_3}^2 + m_p^2} - \sqrt{{\bf p_4^2} + m_p^2} - \sqrt{({\bf p_3} + {\bf p_4})^2 + m_5^2}\right)}{\sqrt{{\bf p_3}^2 + m_p^2}\sqrt{{\bf p_4}^2 + m_p^2}\sqrt{({\bf p_3} + {\bf p_4})^2 + m_5^2}} \, \frac{d^3{\bf p_3}}{(2\pi)^3} \, \frac{d^3{\bf p_4}}{(2\pi)^3}.
\eeq
Now we can rewrite the integrals over ${\bf p_3}$ and ${\bf p_4}$, using our decomposition above, as
\beq
d^3{\bf p_3} \, d^3{\bf p_4} = q_3 \, dq_3 \, d\theta_3 \, dp_{z3} \times q_4 \, dq_4 \, d\theta_4 \, dp_{z4} = \frac{1}{4}d(q_3^2) \, d(q_4^2) \, d\theta_3 \, d\theta_4 \, dp_{3z} \, dp_{4z}.
\eeq
With the definitions
\beq
\theta_3 = \phi, \hspace{.5in} \theta_4 = \phi + \theta_{34}, \hspace{.5in} p_{+z} = p_{3z} + p_{4z} = p \, x_F, \hspace{.5in} p_{-z} = p_{3z} - p_{4z},
\eeq
we obtain
\beq
d\theta_3 \, d\theta_4 = d\phi \, d\theta_{34}, \hspace{.5in} dp_{3z} \, dp_{4z} = \frac{1}{2} \, dp_{+z} \, dp_{-z} = \frac{p}{2} \, dx_F \, dp_{-z}.
\eeq
The integrals over $\theta_{34}$ and $\phi$ are each carried out over the region $[0, 2\pi]$, the integral over $p_{-z}$ will be carried out over $[-\infty, +\infty]$, and the integral over $x_F$ will be carried out over $[-1, +1]$.  Furthermore, we expect the amplitude to have azimuthal symmetry, and thus be independent of $\phi$, so we can carry that integral out explicitly.  Putting these pieces together gives us the total cross section
\beq
\sigma = \frac{p}{2^{9}(2 \pi)^{4}\sqrt{(p_1 \cdot p_2)^2 - m_p^4}}
\eeq
$$
\times \int \frac{\langle |\mathcal{A}|^2\rangle \delta\left(2E - \sqrt{{\bf p_3}^2 + m_p^2} - \sqrt{{\bf p_4^2} + m_p^2} - \sqrt{({\bf p_3} + {\bf p_4})^2 + m_5^2}\right)}{\sqrt{{\bf p_3}^2 + m_p^2}\sqrt{{\bf p_4}^2 + m_p^2}\sqrt{({\bf p_3} + {\bf p_4})^2 + m_5^2}} \, d\theta_{34} \, dx_F \, d(q_3^2) \, d(q_4^2) \, dp_{-z}.
$$
We then use the remaining delta function to perform the integral over $p_{-z}$, giving
\beq
\sigma = \frac{p}{2^{8}(2 \pi)^{4}\sqrt{(p_1 \cdot p_2)^2 - m_p^4}} \int \frac{\langle |\mathcal{A}|^2\rangle}{E_5|p_{3z}E_4 - p_{4z}E_3|} \, d\theta_{34} \, dx_F \, d(q_3^2) \, d(q_4^2)
\eeq
where the kinematic parameters $\{p_{3z}, p_{4z}, E_3, E_4, E_5\}$ are now understood to be expressed in terms of $\{q_3, q_4, x_F, \theta_{34}\}$, using the mass shell and 4-momentum conservation equations.  Note that so far we have made no use of the Regge limit.  We should also remember that we eventually want to work with the process in terms of $\{t_1, t_2, x_F,\theta_{34}\}$, which means we want to rewrite the integrals over $q_3^2$ and $q_4^2$ as integrals over $t_1$ and $t_2$.  However, it will be much easier to understand how to do this once we work out the Regge limit.

\subsection{Regge Limit}

First we note that, in terms of Mandelstam variables, we have
\beq
p = \frac{1}{2}\sqrt{s - 4m_p^2}, \hspace{.75in} p_{3z} = \frac{s - s_2 + 2t_1 - 3m_p^2}{2\sqrt{s - 4m_p^2}}, \hspace{.75in} p_{4z} = \frac{-s + s_1 - 2t_2 - 3m_p^2}{2\sqrt{s - 4m_p^2}},
\eeq
which means
\beq
x_F = \frac{s_1 - s_2 + 2t_1 - 2t_2}{s - 4m_p^2}.
\eeq
In the Regge limit, we want to have $s, s_1, s_2 \gg \mu, t_1, t_2, m^2$, where $m$ is any of the masses involved, and $\mu = \frac{s_1 s_2}{s}$ is held fixed in the limit.  This implies that in the Regge limit we have
\beq
s x_F \approx s_1 - s_2.
\eeq
We can then rewrite $s_1$ and $s_2$ in terms of $\mu$ and $s x_F$, as
\beq
s_1 \approx \frac{1}{2}\left[s x_F + \sqrt{s^2 x_F^2 + 4s\mu}\right], \hspace{1in} s_2 \approx \frac{1}{2}\left[-s x_F + \sqrt{s^2 x_F^2 + 4s\mu}\right].
\eeq
Using these expressions then gives us
\beq
E_5 = \frac{s_1 + s_2 - 2m_p^2}{2\sqrt{s}} \approx \frac{\sqrt{s^2 x_F^2 + 4s\mu} - 2m_p^2}{2\sqrt{s}} \approx \frac{\sqrt{s}}{2}\sqrt{ x_F^2 + \frac{4\mu}{s}}
\eeq
The appearance of $\frac{1}{E_5}$ in the phase space integral then implies that in the extreme Regge limit $s \rightarrow \infty$ there is a pole at $x_F = 0$, but for any finite $s$ there is just a sharp peak, with $E_5 \approx \sqrt{\mu}$ at $x_F = 0$.  Furthermore, we should have
\beq
q_3^2 \approx -\frac{t_1}{2}\left(2 + x_F - \sqrt{x_F^2 + \frac{4\mu}{s}}\right) - \frac{m_p^2}{4}\left(x_F - \sqrt{x_F^2 + \frac{4\mu}{s}}\right)^2
\eeq
and similarly
\beq
q_4^2 \approx -\frac{t_2}{2}\left(2 - x_F - \sqrt{x_F^2 + \frac{4\mu}{s}}\right) - \frac{m_p^2}{4}\left(x_F + \sqrt{x_F^2 + \frac{4\mu}{s}}\right)^2.
\eeq
Now, in performing the variable transformation necessary to rewrite the integrals over $q_3^2$ and $q_4^2$ in terms of $t_1$ and $t_2$, we should note that $\mu$ will depend on $t_1$ and $t_2$.  However, to leading order in the Regge limit, we will get
\beq
d(q_3^2) \, d(q_4^2) \approx \left[\frac{1}{4}\Big(2 - x_F - |x_F|\Big)\Big(2 + x_F - |x_F|\Big) \ + \ \mathcal{O}\left(\sqrt{\mu/s}\right)\right] \, dt_1 \, dt_2 \approx \left[\Big(1 - |x_F|\Big) \ + \ \mathcal{O}\left(\sqrt{\mu/s}\right) \right] \, dt_1 \, dt_2.
\eeq
We also have
\beq
E_3 = \frac{s - s_2 + m_p^2}{2\sqrt{s}} \approx \frac{s - s_2}{2\sqrt{s}}, \hspace{1in} E_4 = \frac{s - s_1 + m_p^2}{2\sqrt{s}} \approx \frac{s - s_1}{2\sqrt{s}},
\eeq
and this implies that
\beq
|E_3 p_{4z} - E_4 p_{3z}| \approx \frac{s}{2}\left[\Big(1 - |x_F|\Big) \ + \ \mathcal{O}\left(\sqrt{\mu/s}\right)\right].
\eeq
This then demonstrates that in the Regge limit we simply have
\beq
\frac{d(q_3^2) \, d(q_4^2)}{|E_3 p_{4z} - E_4 p_{3z}|} \approx \frac{2}{s} \, dt_1 \, dt_2.
\eeq
(Notice that there is neither a formal pole nor a sharp peak at $|x_F| = 1$.)  This allows us to write our total cross section as
\beq
\sigma \approx \frac{1}{2^6 (2\pi)^4 s^2} \int \frac{\langle |\mathcal{A}|^2\rangle}{\sqrt{x_F^2 + \frac{4\mu}{s}}} \, dx_F \, d\theta_{34} \, dt_1 \, dt_2.
\eeq
Finally, since the phase space is sharply peaked in the far Regge limit around $x_F = 0$, we can approximate this further by replacing $\langle |\mathcal{A}|^2\rangle$ with its value at $x_F = 0$, and evaluating the integral over $x_F$ explicitly.  This gives
\beq
\int_{-1}^{1} \frac{dx_F}{\sqrt{x_F^2 + \frac{4\mu}{s}}} = 2\int_{0}^{\sqrt{\frac{s}{4\mu}}} \, \frac{dx}{\sqrt{1 + x^2}} = 2\ln \left(\sqrt{\frac{s}{4\mu}} + \sqrt{\frac{s}{4\mu} + 1}\right) \approx \ln \left(\frac{s}{\mu}\right)
\eeq
so that
\beq
\sigma \approx \frac{1}{4(4\pi)^4 s^2} \int \langle |\mathcal{A}|^2\rangle \, \ln \left(\frac{s}{\mu}\right) \, d\theta_{34} \, dt_1 \, dt_2.
\eeq

It is also useful to write out what various frame dependent and frame independent quantities will be when expressed in the Regge limit with $x_F = 0$, as these will appear in $\langle |\mathcal{A}|^2\rangle$ and the Reggeization of the propagators.  Note first that we will now have
\beq
s_1 \approx s_2 \approx \sqrt{s\mu}, \hspace{1in} q_3 \approx \sqrt{-t_1}, \hspace{1in} q_4 \approx \sqrt{-t_2}.
\eeq
We can then use the mass-shell condition for the centrally produced meson to determine $\mu$, as
\beq
\mu \approx m_5^2 - t_1 - t_2 + 2\sqrt{t_1 t_2} \, \cos\theta_{34}.
\eeq



\end{appendix}



\end{document}